\documentclass[conference]{IEEEtran}
\usepackage{cite}
\usepackage{amsmath,amssymb,amsfonts}
\usepackage{algorithmic}
\usepackage{graphicx}
\usepackage{textcomp}
\usepackage{xcolor}
\def\BibTeX{{\rm B\kern-.05em{\sc i\kern-.025em b}\kern-.08em
    T\kern-.1667em\lower.7ex\hbox{E}\kern-.125emX}}

\usepackage{subfigure}
\usepackage{graphicx}
\usepackage{amsmath}
\usepackage{bm}
\usepackage{url}
\usepackage{stmaryrd}
\usepackage{balance}
\usepackage{amssymb}
\usepackage{pgfplots}
\usepackage{pifont}
\usepackage{epsfig}
\usepackage{booktabs}
\usepackage{pgffor}
\usepackage{tikz}
\usepackage{multirow}
\usepackage{amsmath}
\usepackage{tikz}
\usetikzlibrary{positioning,arrows,calc}
\usepackage[all]{xy}
\usepackage{makecell}
\usepackage{enumitem}

\makeatletter
\newcommand\xleftrightarrow[2][]{%
	\ext@arrow 9999{\longleftrightarrowfill@}{#1}{#2}}
\newcommand\longleftrightarrowfill@{%
	\arrowfill@\leftarrow\relbar\rightarrow}
\makeatother

\newcommand{\mb}{\mathbf}

\newcommand{\Rho}{\mathrm{P}}

\newcommand{\problem}{\textsc{Anna}}

\newcommand{\our}{\textsl{ActiveIter}}
\newcommand{\ourtwentyround}{\textsl{ActiveIter-100}}
\newcommand{\ourtenround}{\textsl{ActiveIter-50}}
\newcommand{\activerandom}{\textsl{ActiveIter-Rand-50}}
\newcommand{\activerand}{\textsl{ActiveIter-Rand}}
\newcommand{\pusvm}{\textsl{Iter-MPMD}}
\newcommand{\svmmp}{\textsc{SVM-MP}}

\newcommand{\svmmpmd}{\textsc{SVM-MPMD}}

\usepackage{algorithm}
\usepackage{algorithmic}

\begin{document}

\title{Meta Diagram based Active Social Networks Alignment}

\author{\IEEEauthorblockN{Yuxiang Ren}
\IEEEauthorblockA{\textit{IFM Lab} \\
\textit{Florida State University}\\
Tallahassee, USA \\
yuxiang@ifmlab.org}
\and
\IEEEauthorblockN{Charu C. Aggarwal}
\IEEEauthorblockA{\textit{IBM Research AI} \\
	New York, USA \\
	charu@us.ibm.com}
\and
\IEEEauthorblockN{Jiawei Zhang}
\IEEEauthorblockA{\textit{IFM Lab} \\
	\textit{Florida State University}\\
	Tallahassee, USA \\
	jiawei@ifmlab.org}
}
\maketitle

\begin{abstract}
	
	Network alignment aims at inferring a set of anchor links matching the shared entities between different information networks, which has become a prerequisite step for effective fusion of multiple information networks. In this paper, we will study the network alignment problem to fuse online social networks specifically. Social network alignment is extremely challenging to address due to several reasons, i.e., \textit{lack of training data}, \textit{network heterogeneity} and \textit{one-to-one constraint}. Existing network alignment works usually require a large number of training instances, but such a demand can hardly be met in applications, as manual anchor link labeling is extremely expensive. Significantly different from other homogeneous network alignment works, information in online social networks is usually of heterogeneous categories, the incorporation of which in model building is not an easy task. Furthermore, the \textit{one-to-one} cardinality constraint on anchor links renders their inference process intertwistingly correlated. To resolve these three challenges, a novel network alignment model, namely {\our}(Active Iterative Alignment), is introduced in this paper. The model {\our} defines a set of \textit{inter-network meta diagrams} for anchor link feature extraction, adopts \textit{active learning} for effective label query and uses \textit{greedy link selection} for anchor link cardinality filtering. Extensive experiments were performed on a real-world aligned networks dataset, and the experimental results have demonstrated the effectiveness of {\our} compared with other state-of-the-art baseline methods.

\end{abstract}

\section{Introduction}\label{sec:introduction}

\begin{figure*}[t]
	\vspace{-20pt}
	\centering
	\begin{minipage}[l]{1.9\columnwidth}
		\centering
		\includegraphics[width=\textwidth]{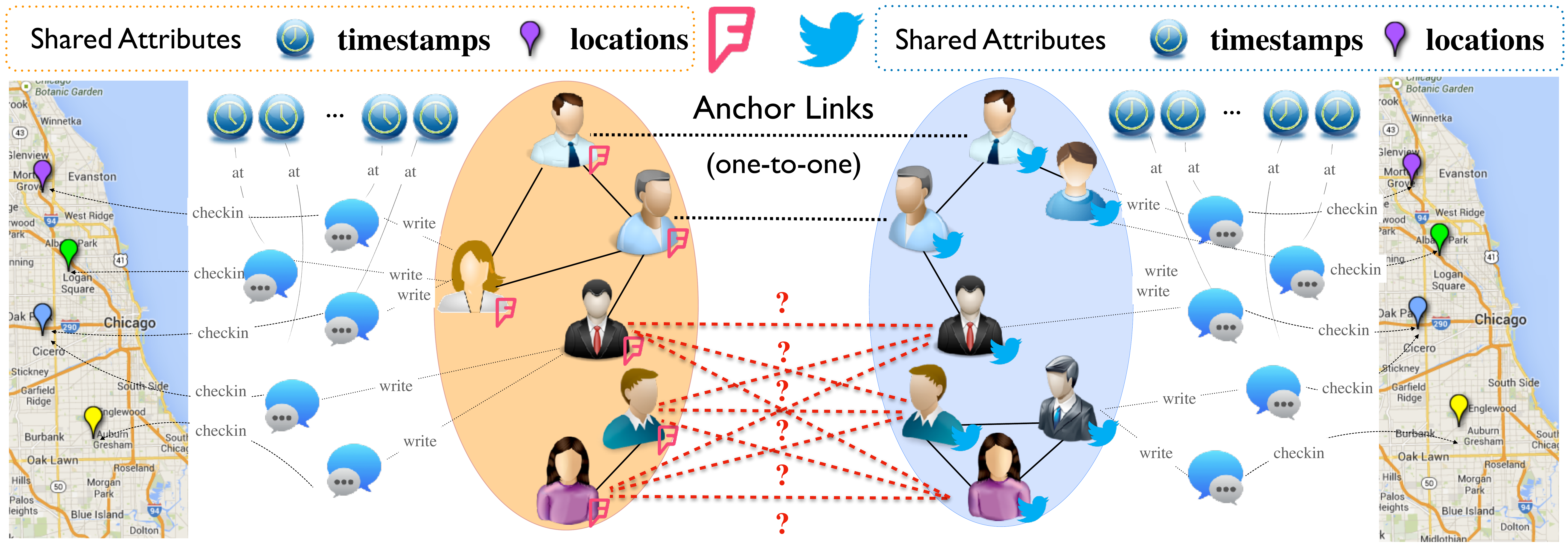}
	\end{minipage}
	\vspace{-10pt}
	\caption{An example of attributed heterogeneous social networks alignment (Foursquare and Twitter).}\label{fig:example}
	\vspace{-15pt}
\end{figure*}

Formally, the network alignment problem \cite{ZY15, FNSMB06} denotes the task of inferring the set of anchor links \cite{KZY13} between the shared information entities in different networks, where the anchor links are usually assumed to be subject to the \textit{one-to-one} cardinality constraint \cite{ZCZCY17}. Network alignment has concrete applications in the real world, which can be applied to discover the set of shared users between different online social networks \cite{ZY15, KZY13}, identify the common protein molecules between different protein-protein-interaction (PPI) networks \cite{SXB07, FNSMB06, SXB08}, and find the mappings of POIs (points of interest) across different traffic networks \cite{ZY15}. In this paper, we will use online social networks as an example of a real world setting of the network alignment problem and also use this setting to elucidate the proposed model.

Online social networks usually have very complex structures, involving different categories of nodes and links. For instance, in online social networks, like Twitter and Foursquare as shown in Figure~\ref{fig:example}, users can perform various kinds of social activities, e.g., following other users, writing posts. Viewed in such a perspective, their network structures will contain multiple types of nodes and links, i.e., ``User'', ``Post'' (node types), and ``Follow'', ``Write'' (link types). Users' personal preference may steer their online social activities, and the network structure can provide insightful information for differentiating users between networks. Furthermore, the nodes in online social networks can be also attached with various types of attributes. For example, the written post nodes can contain words, location check-ins and timestamps (attribute types), which will provide complementary information for inferring users' language usage, spatial and temporal activity patterns respectively. Based on such an intuition, both the network structure and attribute information should be incorporated in the network alignment model building.

Most of the existing network alignment models are based on supervised learning \cite{KZY13}, which aim at building classification/regression models with a large set of pre-labeled anchor links to infer the remaining unlabeled ones (where the existing and non-existing anchor links are labeled as the positive and negative instance respectively). For the network alignment task, pre-labeled anchor links can provide necessary information for understanding the patterns of aligned user pairs in their information distribution, especially compared with the unsupervised alignment models \cite{ZY15, FNSMB06}. However, for the real-world online social networks, cross-network anchor link labeling is not an easy task, since it requires tedious user-account pairing and manual user-background checking, which can be very time-consuming and expensive. Therefore, a large training data set as required by existing network alignment models \cite{KZY13} is rarely available in the real world.

\noindent \textbf{Problem Studied}: In this paper, we propose to study the heterogeneous network alignment problem based on the active learning setting, which is formally referred to the \textit{\underline{A}ctive heteroge\underline{N}eous \underline{N}etwork \underline{A}lignment} ({\problem}) problem. Subject to the pre-specified query budget (i.e., the label query times), {\problem} allows the models to selectively query for extra labels of the unlabeled anchor links in the learning process. In Figure~\ref{fig:example}, we shown an example of the {\problem} problem between the Foursquare and Twitter social networks.

The current research has not studied the heterogeneous network alignment problem based on active learning setting yet. The {\problem} problem is a novel yet difficult task, and the challenges mainly come from three perspectives, e.g., \textit{network heterogeneity}, \textit{paucity of training data}, and \textit{one-to-one constraint}. 

\begin{itemize}
	
	\item \textit{Network Heterogeneity}: According to the aforementioned descriptions, both the complex network structure and the diverse attributes have concrete physical meanings and can be useful for the social network alignment task. To incorporate such heterogeneous information in model building, a unified approach is required to handle the network structure and attribute information in a unified analytic.
	
	\item \textit{Paucity of Training Data}: To overcome problems caused by paucity of training data, besides the labeled anchor links, active learning also allows models to query for extra labels of unlabeled instances. In this context, active learning application in network alignment still remains unexplored. 
	
	
	\item \textit{One-to-One Cardinality Constraint}: Last but not the least, the anchor links to be inferred are not independent in the networked data scenario. The \textit{one-to-one} cardinality constraint on anchor links will limit the number of anchor links incident to the user nodes \cite{ZCZCY17,KZY13}, which renders the information of positive and negative anchor links to be imbalanced. For each user, if one incident anchor link is identified to be positive, the remaining incident anchor links will all be negative by default. Viewed from such a perspective, positive anchor links contribute far more information compared with the negative ones. Effectively maintaining and utilizing such a constraint on anchor links in the active label query and model building is a challenging problem.
	
\end{itemize}

To address these challenges, we will introduce a new network alignment model, namely \textit{Active Iterative Alignment} ({\our}), in this paper. To model the diverse information available in social networks, {\our} adopts the \textit{attributed heterogeneous social network} concept to represent the complex network structure and the diverse attributes on nodes and links. Furthermore, a unified feature extraction method will be introduced in {\our}, based on a novel concept namely \textit{meta diagram}, for anchor links between attributed heterogeneous social networks. {\our} accepts coupled user pairs as the input, and outputs the inference results of the anchor links between them utilizing information about both the labeled and unlabeled anchor links. To deal with the paucity of training data, active learning will be adopted in {\our} to utilize the unlabeled anchor links in model building by querying for extra anchor link labels based on a designated stratedy within certain pre-specified query budget. Due to the \textit{one-to-one} constraint, the unlabeled anchor links no longer bears equal information, and querying for labels of potential positive anchor links will be more ``informative'' compared with negative anchor links. Among the unlabeled links, {\our} aims at selecting a set of mis-classified false-negative anchor links as the potential candidates. Using such an approach contributes to not only these queried labels but also other potential extra label corrections of the conflicting negative links.
An innovative query strategy is proposed to make sure that {\our} can select mis-classified false-negative anchor links more precisely. {\our} can outperform other non-active models with less than $10\%$ of extra training instances which has the additional benefits of reducing the time and space complexity. 

The remaining parts of this paper will be organized as follows. In Section~\ref{sec:formulation}, we will introduce the definitions of several important terminologies and the formal problem statement. Detailed information about the proposed model will be provided in Section~\ref{sec:method}, whose effectiveness and efficiency will be tested in Section~\ref{sec:experiment}. Related works will be talked about in Section~\ref{sec:related_work}, and finally in Section~\ref{sec:conclusion} we will conclude this paper.

\section{Concept and Problem Definition} \label{sec:formulation}

In this section, we will define several important concepts used in this paper, and provide the formulation of the {\problem} problem.

\subsection{Terminology Definition}

\noindent \textbf{Definition 1} (Attributed Heterogeneous Social Networks): The \textit{attributed heterogeneous social network} studied in this paper can be represented as $G = (\mathcal{V}, \mathcal{E}, \mathcal{T})$, where $\mathcal{V} = \bigcup_i \mathcal{V}_i$ and $\mathcal{E} = \bigcup_i \mathcal{E}_i$ represent the sets of diverse nodes and complex links in the network. The set of attributes associated with nodes in $\mathcal{V}$ can be represented as set $\mathcal{T}$ = $\bigcup_i \mathcal{T}_i$ ($\mathcal{T}_i$ denotes the $i_{th}$-type of attributes).


Meanwhile, for the \textit{attributed heterogeneous social networks} with shared users, they can be represented as the multiple \textit{aligned attributed heterogeneous social networks} (or \textit{aligned social networks} for short).


\noindent \textbf{Definition 2} (Multiple Aligned Social Networks): Given online social networks $G^{(1)}$, $G^{(2)}$, $\cdots$, $G^{(n)}$ sharing common users, they can be represented as the \textit{multiple aligned social networks} $\mathcal{G} = \left( (G^{(1)}, G^{(2)}, \cdots, G^{(n)}), (\mathcal{A}^{(1,2)}, \mathcal{A}^{(1,3)}, \cdots, \\ \mathcal{A}^{(n-1, n)}) \right)$, where $\mathcal{A}^{(i,j)}$ represents the set of undirected anchor links connecting the common users between networks $G^{(i)}$ and $G^{(j)}$.


In Figure~\ref{fig:example}, we show an example of two \textit{aligned social networks}, Foursquare and Twitter, which can be represented as $\mathcal{G} = ((G^{(1)}, G^{(2)}), \mathcal{A}^{(1,2)})$. Formally, the Twitter network can be represented as $G^{(1)} = (\mathcal{V}^{(1)}, \mathcal{E}^{(1)}, \mathcal{T}^{(1)})$, where $\mathcal{V}^{(1)} = \mathcal{U}^{(1)} \cup \mathcal{P}^{(1)}$ denotes the set of nodes in the network including users and posts, and $\mathcal{E}^{(1)} = \mathcal{E}_{u,u}^{(1)} \cup \mathcal{E}_{u,p}^{(1)}$ involves the sets of social links among users as well as write links between users and posts. For the posts, a set of attributes can be extracted, which can be represented as $\mathcal{T}^{(1)} = \mathcal{T}^{(1)}_{w} \cup \mathcal{T}^{(1)}_{l} \cup \mathcal{T}^{(1)}_{t}$ denoting the words, location checkins and timestamps attached to the posts in $\mathcal{P}^{(1)}$ respectively. The Foursquare network has a similar structure as Twitter, which can be represented as $G^{(2)} = (\mathcal{V}^{(2)}, \mathcal{E}^{(2)}, \mathcal{T}^{(2)})$. Twitter and Foursquare are aligned together by the user anchor links connecting the shared users, and they also share some common attributes at the same time.

In this paper, we will use these two \textit{aligned social networks} $\mathcal{G} = ((G^{(1)}, G^{(2)}), \mathcal{A}^{(1,2)})$ as an example to illustrate the problem setting and proposed model, but simple extensions of the model can be applied to multiple (more than two) aligned social networks as well.

\subsection{Problem Definition}\label{subsec:problem_definition}

\noindent \textbf{Problem Definition}: Given a pair of partially \textit{aligned social networks} $\mathcal{G} = ((G^{(1)}, G^{(2)}), \mathcal{A}^{(1,2)})$, we can represent all the potential anchor links between networks $G^{(1)}$ and $G^{(2)}$ as set $\mathcal{H} = \mathcal{U}^{(1)} \times \mathcal{U}^{(2)}$, where $\mathcal{U}^{(1)}$ and $\mathcal{U}^{(2)}$ denote the user sets in $G^{(1)}$ and $G^{(2)}$ respectively. For the known links between networks, we can group them as a labeled set  $\mathcal{L} = \mathcal{A}^{(1,2)}$. The remaining anchor links with unknown labels are those to be inferred, and they can be formally denoted as the unlabeled set $\mathcal{U} = \mathcal{H} \setminus \mathcal{L} $. In the {\problem} problem, based on both labeled anchor links in $\mathcal{L}$ and unlabeled anchor links in $\mathcal{U}$, we aim at building a mapping function $f: \mathcal{H} \to \mathcal{Y}$ to infer anchor link labels in $\mathcal{Y} = \{0, +1\}$ subject to the \textit{one-to-one} constraint, where class labels $+1$ and $0$ denote the existing and non-existing anchor links respectively. Besides these known links, in the {\problem} problem, we are also allowed to query for the label of links in set $\mathcal{U}$ with a pre-specified budget $b$, i.e., the number of allowed queries. Besides learning the optimal variables in the mapping function $f(\cdot)$, we also aim at selecting an optimal query set $\mathcal{U}_q$ to improve the performance of the learned mapping function $f(\cdot)$ as much as possible.


\section{Proposed Method}\label{sec:method}

\begin{figure}[t]
	\vspace{-10pt}
	\centering
	\begin{minipage}[l]{1.0\columnwidth}
		\centering
		\includegraphics[width=\textwidth]{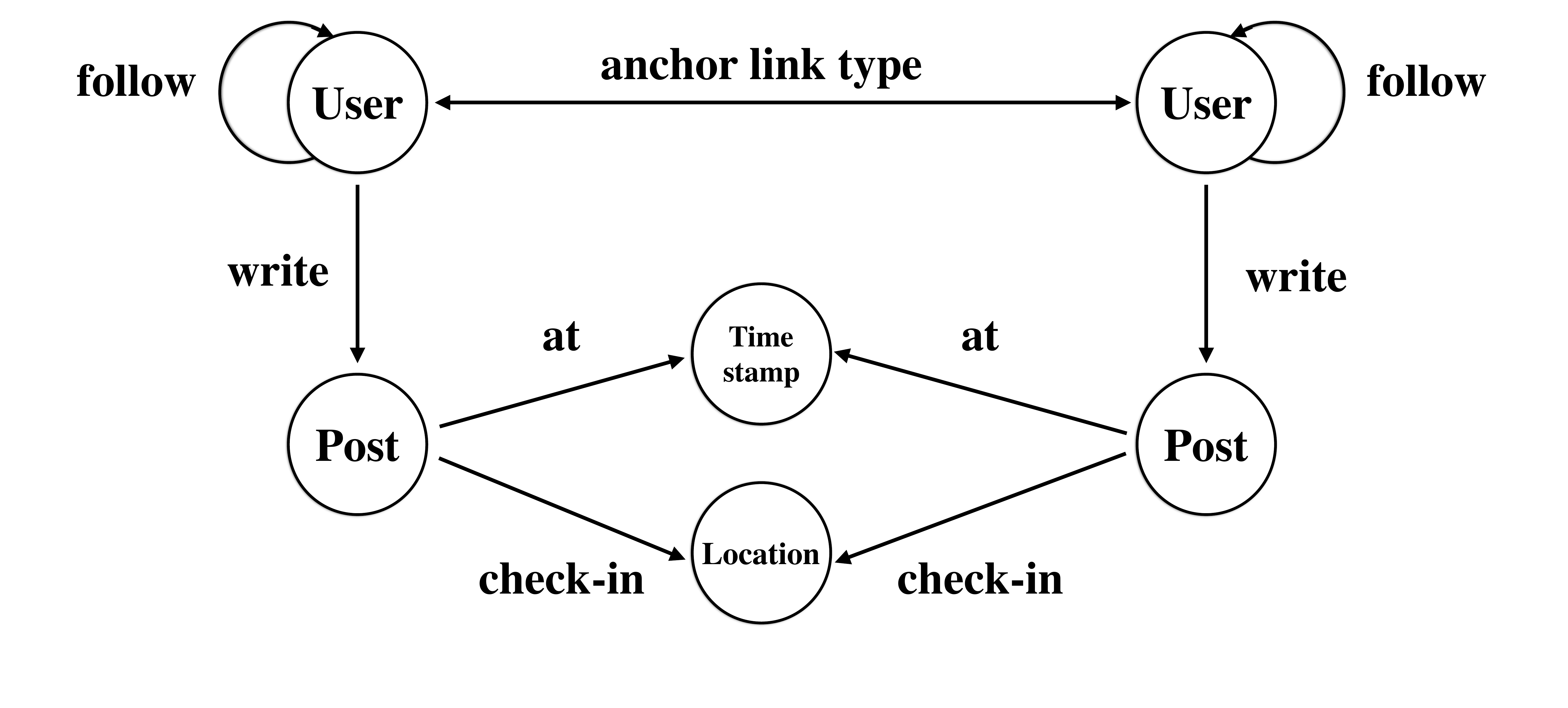}
	\end{minipage}
	\vspace{-12pt}
	\caption{Schema of aligned networks.}\label{fig:schema}
	\vspace{-12pt}
\end{figure}

In this section, we will introduce the proposed model {\our} in detail. At the very beginning, we will introduce the notations used in this paper. After that, the formal definition of \textit{Meta Diagram} will be provided, based on which a set of meta diagram based features will be extracted. Finally, we will introduce the active network alignment model for anchor link inference.


\subsection{Notations}\label{subsec:notation}

In the sequel, we will use lower case letters (e.g., $x$) to denote scalars, lower case bold letters (e.g., $\mb{x}$) to denote column vectors, bold-face upper case letters (e.g., $\mb{X}$) to denote matrices, and upper case calligraphic letters (e.g., $\mathcal{X}$) to denote sets. The $i_{th}$ entry of vector $\mb{x}$ is denoted as $x(i)$. Given a matrix $\mb{X}$, we denote $\mb{X}(i,:)$ (and $\mb{X}(:,j)$) as the $i_{th}$ row (and the $j_{th}$ column) of $\mb{X}$, and the $(i_{th}, j_{th})$ entry of matrix $\mb{X}$ can be denoted as $X(i,j)$ or $X_{i,j}$ (which are interchangeable). We use $\mb{X}^\top$ (and $\mb{x}^\top$) to denote the transpose of matrix $\mb{X}$ (and vector $\mb{x}$). For vector $\mb{x}$, we denote its $L_p$-norm as $\left \| \mb{x} \right \|_p = (\sum_i |x_i|^p)^{\frac{1}{p}}$, and the $L_p$-norm of matrix $\mb{X}$ can be represented as $\left \| \mb{X} \right \|_p = (\sum_{i,j} |X(i,j)|^p)^{\frac{1}{p}}$. Given two vectors $\mb{x}$, $\mb{y}$ of the same dimension, we use notation $\mb{x} \le \mb{y}$ to denote that entries in $\mb{x}$ are no greater than the corresponding entries in $\mb{y}$.


\begin{table*}[t]

	\scriptsize
	\vspace{-10pt}
	\centering
	{
		\vspace{-10pt}
		\caption{Summary of Inter-Network Meta Diagrams.}\label{tab:meta_path}
		\vspace{-5pt}
		\begin{tabular}{llll}
			\hline
			\textbf{ID}
			&\textbf{Notation}
			& \textbf{Meta Diagram}
			& \textbf{Semantics}\\
			\hline
			
			$\Rho_1$
			&U $\to$ U $\leftrightarrow$ U $\gets$ U
			&User $\xrightarrow{follow}$ User $\xleftrightarrow{anchor}$ User $\xleftarrow{follow}$ User
			&Common Anchored Followee\\
			
			$\Rho_2$
			&U $\gets$ U $\leftrightarrow$ U $\to$ U
			&User $\xleftarrow{follow}$ User $\xleftrightarrow{anchor}$ User $\xrightarrow{follow}$ User
			&Common Anchored Follower\\
			
			$\Rho_3$
			&U $\to$ U $\leftrightarrow$ U $\to$ U
			&User $\xrightarrow{follow}$ User $\xleftrightarrow{anchor}$ User $\xrightarrow{follow}$ User
			&Common Anchored Followee-Follower\\
			
			$\Rho_4$
			&U $\gets$ U $\leftrightarrow$ U $\gets$ U
			&User $\xleftarrow{follow}$ User $\xleftrightarrow{anchor}$ User $\xleftarrow{follow}$ User
			&Common Anchored Follower-Followee\\
			
			
			$\Rho_5$
			&U $\to$ P $\to$ T $\gets$ P $\gets$ U
			&User $\xrightarrow{write}$ Post $\xrightarrow{at}$ Timestamp $\xleftarrow{at}$ Post $\xleftarrow{write}$ User
			&Common Timestamp\\
			
			$\Rho_6$
			&U $\to$ P $\to$ L $\gets$ P $\gets$ U
			&User $\xrightarrow{write}$ Post $\xrightarrow{checkin}$ Location $\xleftarrow{checkin}$ Post $\xleftarrow{write}$ User
			&Common Checkin\\
			
			\hline
			
			$\Psi_1 (\Rho_1 \times \Rho_2)$
			
			&U $\leftrightarrow$ U $\xleftrightarrow{anchor}$ U $\leftrightarrow$ U
			
			&
			\begin{tikzpicture}[baseline={([yshift=-.5ex]current bounding box.center)}, node distance=1cm, auto,]
			\node[] (dummy) {};
			\coordinate (CENTER) at ($(dummy)$);
			
			\node[right=of dummy] (post-right) {User};
			
			\node[left=of dummy] (post-left) {User}
			edge[<->] node[midway,above] {{\tiny $anchor$}} (post-right.west);

			\node[above=0.01cm of post-right] (user-right-above) {};
			\node[above=0.01cm of post-left] (user-left-above) {};
			
			\node[below=0.01cm of post-right] (user-right-below) {};
			\node[below=0.01cm of post-left] (user-left-below) {};
			
			\node[left=of post-left] (user-left) {User}
			edge[->, bend left=10] node[midway,above] {{\tiny $follow$}} (user-left-above.west)
			edge[<-, bend left=-10] node[midway,below] {{\tiny $follow$}} (user-left-below.west);
			
			\node[right=of post-right] (user-right) {User}
			edge[->, bend left=-10] node[midway,above] {{\tiny $follow$}} (user-right-above.east)
			edge[<-, bend left=10] node[midway,below] {{\tiny $follow$}} (user-right-below.east);
			
			\end{tikzpicture}
			&\makecell[l]{Common Aligned Neighbors}\\

			$\Psi_2 (\Rho_5 \times \Rho_6)$
			&
			\begin{tikzpicture}[baseline={([yshift=-.5ex]current bounding box.center)}, node distance=1cm]
			\node (U) at (-.15, 0) {U};
			\node (P) at (0.5, 0) {P};
			\node (T) at (1.15, -0.3) {T};
			\node (L) at (1.15, 0.3) {L};
			\node (P2) at (1.8, 0) {P};
			\node (U2) at (2.45, 0) {U};
			
			\draw[->, to path={|- (\tikztotarget)}]
			(U) edge (P) (P) edge (L) (P) edge (T);
			\draw[<-, to path={-| (\tikztotarget)}]
			(L) edge (P2) (T) edge (P2);
			\draw[<-, to path={-| (\tikztotarget)}]
			(P2) edge (U2.west);
			
			\end{tikzpicture}&User $\xrightarrow{write}$  
			\begin{tikzpicture}[baseline={([yshift=-.5ex]current bounding box.center)}, node distance=1cm, auto,]
			\node[] (dummy) {};
			\coordinate (CENTER) at ($(dummy)$);
			\node[inner sep=0pt,above=0.0cm of dummy] (location) {Location};
			\node[inner sep=0pt,below=0.0cm of dummy] (time) {Timestamp};
			
			\node[right=of dummy] (post-right) {Post}
			edge[->, bend right=10] node[midway,above] {{\tiny $checkin$}} (location.east) 
			edge[->, bend left=10] node[midway,below] {{\tiny $at$}} (time.east); 
			
			\node[left=of dummy] (post-left) {Post}
			edge[->, bend left=10] node[midway,above] {{\tiny $checkin$}} (location.west)
			edge[->, bend right=10] node[midway,below] {{\tiny $at$}} (time.west);
			\end{tikzpicture}
			$\xleftarrow{write}$ User
			&\makecell[l]{Common Attributes}\\
			
			$\Psi_3 (\Rho_1 \times \Rho_5 \times \Rho_6)$
			&
			\begin{tikzpicture}[baseline={([yshift=-.5ex]current bounding box.center)}, node distance=1cm]
			\node (U11) at (-.15, 0.5) {U};
			\node (U) at (-.15, 0) {U};
			\node (P) at (0.5, 0) {P};
			\node (W) at (1.15, -0.2) {T};
			\node (L) at (1.15, 0.2) {L};
			\node (P2) at (1.8, 0) {P};
			\node (U2) at (2.45, 0) {U};
			\node (U22) at (2.45, 0.5) {U};
			
			\draw[->, to path={|- (\tikztotarget)}]
			(U) edge (P) (P) edge (L) (P) edge (W) (U) edge (U11.south) (U2) edge (U22.south);
			\draw[<-, to path={-| (\tikztotarget)}]
			(L) edge (P2) (W) edge (P2);
			\draw[<-, to path={-| (\tikztotarget)}]
			(P2) edge (U2.west);
			\draw[<-, to path={-| (\tikztotarget)}]
			(U11) edge (U22.west);
			\draw[->, to path={|- (\tikztotarget)}]
			(U11) edge (U22.west);
			\end{tikzpicture}
			
			&
			\begin{tikzpicture}[baseline={([yshift=-.5ex]current bounding box.center)}, node distance=1cm, auto,]
			\node[] (dummy) {};
			\coordinate (CENTER) at ($(dummy)$);
			\node[inner sep=0pt,above=0.0cm of dummy] (location) {Location};
			\node[inner sep=0pt,below=0.0cm of dummy] (time) {Timestamp};
			
			\node[left=of dummy] (post-left) {Post}
			edge[->, bend left=10] node[midway,above] {{\tiny $checkin$}} (location.west)
			edge[->, bend right=10] node[midway,below] {{\tiny $at$}} (time.west);
			
			\node[right=of dummy] (post-right) {Post}
			edge[->, bend right=10] node[midway,above] {{\tiny $checkin$}} (location.east) 
			edge[->, bend left=10] node[midway,below] {{\tiny $at$}} (time.east); 
			
			\node[above=0.2cm of post-right] (user-right-above) {User};
			\node[above=0.2cm of post-left] (user-left-above) {User}
			edge[<->] node[midway,above] {{\tiny $anchor$}} (user-right-above.west);
			
			\node[left=of post-left] (user-left) {User}
			edge[->] node[midway,above] {{\tiny $write$}} (post-left.west)
			edge[->, bend left=10] node[midway,above] {{\tiny $follow$}} (user-left-above.west);
			
			\node[right=of post-right] (user-right) {User}
			edge[->] node[midway,above] {{\tiny $write$}} (post-right.east)
			edge[->, bend left=-10] node[midway,above] {{\tiny $follow$}} (user-right-above.east);
			
			\end{tikzpicture}
			&\makecell[l]{Common Aligned Neighbor \& Attributes}\\

			\hline
			
		\end{tabular}
	}
	\vspace{-10pt}
\end{table*}

\subsection{Meta Diagram based Proximity Features}

The \textit{attributed heterogeneous social network} introduced in Section~\ref{sec:formulation} provides a unified representation for most of the popular online social networks, like Facebook, Twitter and Foursquare.

\subsubsection{Network Schema and Inter-Network Meta Path}

To effectively categorize the diverse information in the \textit{aligned social networks}, we introduce the \textit{aligned network schema} concept as follows.

\noindent \textbf{Definition 3} (Aligned Social Network Schema): Formally, the schema of the given aligned social networks $\mathcal{G} = ((G^{(1)}, G^{(2)}), \mathcal{A}^{(1,2)})$ can be represented as $S_{\mathcal{G}} = ((S_{{G}^{(1)}}, S_{{G}^{(2)}}), \{\mbox{anchor}\})$. Here, $S_{{G}^{(1)}} = (\mathcal{N}^{(1)}_{\mathcal{V}} \cup \mathcal{N}_{\mathcal{T}}, \mathcal{R}_{\mathcal{E}} \cup \mathcal{R}_{\mathcal{A}})$, where $\mathcal{N}^{(1)}_{\mathcal{V}}$ and $\mathcal{N}_{\mathcal{T}}$ denote the set of node types and attribute types in the network, while  $\mathcal{R}_{\mathcal{E}}$ represents the set of link types in the network, and $\mathcal{R}_{\mathcal{A}}$ denotes the set of association types between nodes and attributes. In a similar way, we can represent the schema of $G^{(2)}$ as $S_{{G}^{(2)}} = (\mathcal{N}^{(2)}_{\mathcal{V}} \cup \mathcal{N}_{\mathcal{T}}, \mathcal{R}_{\mathcal{E}} \cup \mathcal{R}_{\mathcal{A}})$.


In the above definition, to simplify the representations, (1) the attribute types have no superscript, since lots of attribute types can be shared across networks; and (2) the relation types also have no superscript, and the network they belong to can be easily differentiated according to the superscript of user/post node types connected to them. According to the definition, as shown in Figure~\ref{fig:schema}, we can represent the Twitter network schema as $S_{G^{(1)}} = (\mathcal{N}^{(1)}, \mathcal{R})$, $\mathcal{N}^{(1)} =  \{$User$^{(1)}$, Post$^{(1)}$, Word, {Location}, {Timestamp}$\}$ (or $\mathcal{N}^{(1)} =  \{$U$^{(1)}$, P$^{(1)}$, W, {L}, {T}$\}$ for short) and $\mathcal{R} = \{$follow, write, at, check-in$\}$. The Foursquare network schema has exactly the same representation, and it can be denoted as $S_{G^{(2)}} = (\mathcal{N}^{(2)}, \mathcal{R})$, where $\mathcal{N}^{(2)} =  \{$U$^{(2)}$, P$^{(2)}$, W, {L}, {T}$\}$ and $\mathcal{R} = \{$follow, write, at, check-in$\}$. Nodes between Twitter and Foursquare can be connected with each other via connections consisting of various types of links. To categorize all these possible connections across networks, we define the concept of \textit{inter-network meta path} based on the schema as follows:

\noindent \textbf{Definition 4} (Inter-Network Meta Path): Based on an aligned attributed network schema, $S_{\mathcal{G}} = ((S_{{G}^{(1)}}, S_{{G}^{(2)}}), \{\mbox{anchor}\})$, path $\Rho = N_1 \xrightarrow{R_1} N_2 \xrightarrow{R_2} \cdots \xrightarrow{R_{n-1}} N_n$ is defined to be an \textit{inter-network meta path} of length $n-1$ between networks $G^{(1)}$ and $G^{(1)}$, where $N_i \in \mathcal{N}^{(1)} \cup \mathcal{N}^{(2)}, i \in \{1, 2, \cdots, n\}$ and $R_i \in \mathcal{R} \cup \{anchor\}, i \in \{1, 2, \cdots, n-1\}$. In this paper, we are only concerned about \textit{inter-network meta paths} connecting users across networks, in which $N_1, N_n \in \{\mbox{U}^{(1)}, \mbox{U}^{(2)}\} \land N_1 \neq N_n$.

Based on the aligned network schema shown in Figure~\ref{fig:schema}, several \textit{inter-network meta paths} $\{\Rho_1, \Rho_2, \cdots, \Rho_6\}$ can be defined, whose physical meanings and notations are summarized in the top part of Table~\ref{tab:meta_path}.

\subsubsection{Inter-Network Meta Diagram}

For the applications on real-world online social networks, these meta paths extracted in the pervious subsection may suffer from two major disadvantages. Firstly, meta path cannot characterize rich semantics. For instance, given two users $u_i^{(1)}$ and $u_j^{(2)}$ with check-in records ``$u_i^{(1)}$: (Chicago, Aug. 2016), (New York, Jan. 2017), (Los Angeles, May 2017)'', and ``$u_j^{(2)}$: (Los Angeles, Aug. 2016), (Chicago, Jan. 2017), (New York, May 2017)'' respectively, based on meta path $\Rho_5$ and $\Rho_6$, user pair $u_i^{(1)}$, $u_j^{(2)}$ have a lot in common and are highly likely to be the same user, since they have either checked-in the same locations (for $3$ times) or at the same time (for $3$ times). However, according to their check-in records, we observe that their activities are totally ``dislocated'' as they have never been at the same place for the same moments. Secondly, different meta paths denote different types of connections among users, and assembling them in an effective way is another problem. Actually, the meta paths can not only been concatenated but also stacked. Based on such an intuition, to solve these two challenges, we introduce a new concept \textit{Inter-Network Meta Diagram}, which is a meta subgraph that fuses diverse relationships together for capturing richer semantic information across aligned attributed heterogeneous networks specifically. \textit{Inter-network meta diagram} is different from the intra-network \textit{meta graph} \cite{ZYLSL17} and \textit{meta structure} \cite{HZCSML16} concepts proposed in the existing works, since it mainly exists across multiple heterogeneous networks. More detailed information about these concepts and their differences will be provided in Section~\ref{sec:related_work}.

\noindent \textbf{Definition 5} (Inter-Network Meta Diagram): Give a network schema as $S_{\mathcal{G}} = ((S_{{G}^{(1)}}, S_{{G}^{(2)}}), \{\mbox{anchor}\})$, an \textit{inter-network meta diagram} can be formally represented as a directed acyclic subgraph $\Psi = (\mathcal{N}_{\Psi}, \mathcal{R}_{\Psi}, N_s, N_t)$, where $\mathcal{N}_{\Psi} \subset \mathcal{N}^{(1)} \cup \mathcal{N}^{(2)}$ and $\mathcal{R}_{\Psi} \subset \mathcal{R} \cup \{anchor\}$ represents the node, attribute and link types involved, while $N_s, N_t \in \{\mbox{U}^{(1)}, \mbox{U}^{(2)}\} \land N_s \neq N_t$ denote the source and sink user node types from network $G^{(1)}$ and $G^{(2)}$ respectively.


\textit{Inter-network meta diagram} proposed for the \textit{aligned attributed heterogeneous networks} involves not only regular node types but also attribute types and it connects user node types across networks, which renders it different from the recent intra-network \textit{meta structure} \cite{HZCSML16} or \textit{meta graph} \cite{ZYLSL17} concepts proposed for single non-attributed networks. Several \textit{meta diagram} examples have been extracted from the networks as shown at the bottom part of Table~\ref{tab:meta_path} which can be represented as $\{\Psi_1, \Psi_2, \Psi_3\}$. Here, the \textit{meta diagrams} $\Psi_1$ and $\Psi_2$ are composed of $2$ meta paths based on social relationship and anchor (i.e., $\Rho_1$ and $\Rho_2$), as well as attributes (i.e., $\Rho_5$ and $\Rho_6$) respectively; $\Psi_3$ is composed of $3$ meta paths $\Rho_1$, $\Rho_5$ and $\Rho_6$ respectively. Besides these listed meta diagram examples shown in Table~\ref{tab:meta_path}, several other \textit{meta diagrams} are also extracted. Formally, we can use $\Rho_{f} = \{\Rho_1, \Rho_2, \Rho_3, \Rho_4\}$ and $\Rho_{a} = \{\Rho_5, \Rho_6\}$ represent the sets of meta paths composed of the social relationships and the attributes respectively. The complete list of inter-network meta diagrams extracted in this paper are listed as follows:

\noindent $\bullet$ $\Psi_{f^2}$ ($\Rho_{f} \times \Rho_{f}$): Common Aligned Neighbor\underline{\textbf{s}}.

\noindent $\bullet$  $\Psi_{a^2}$ ($\Rho_{a} \times \Rho_{a}$): Common Attribute\underline{\textbf{s}}.

\noindent $\bullet$  $\Psi_{f,a}$ ($\Rho_{f} \times \Rho_{a}$): Common Aligned Neighbor \& Attribute.

\noindent $\bullet$  $\Psi_{f,a^2}$ ($\Rho_{f} \times \Rho_{a} \times \Rho_{a}$): Common Aligned Neighbor \& Attribute\underline{\textbf{s}}.


\noindent $\bullet$  $\Psi_{f^2,a^2}$ ($\Rho_{f} \times \Rho_{f} \times \Rho_{a} \times \Rho_{a}$): Common Aligned Neighbor\underline{\textbf{s}} \& Attribute\underline{\textbf{s}}.


Here, $\Psi_{f^2} = \Rho_{f} \times \Rho_{f} = \{\Rho_i \times \Rho_j\}_{\Rho_i \in \Rho_{f}, \Rho_j \in \Rho_{f}}$, and $\Psi_{f,a} = \Rho_{f} \times \Rho_{a} = \{\Rho_i \times \Rho_j\}_{\Rho_i \in \Rho_{f}, \Rho_j \in \Rho_{a}}$, and similar for the remaining notations. The operator $\Rho_i \times \Rho_j$ denotes the stacking of meta paths $\Rho_i$ and $\Rho_j$ via the common node types shared by them. For instance, $\Psi_1$ is an anchor meta diagram composed by stacking two anchor meta paths of social relationships, i.e., $\Psi_1 \in \Psi_{f^2}$. Actually, \textit{meta path} is also a special type of \textit{meta diagram} in the shape of path. To unify the terms, we will misuse \text{meta diagram} to refer to both \textit{meta path} and \textit{meta diagram} in this paper. Formally, all the \textit{meta diagrams} extracted from the social networks can be represented as $\Phi = \Rho \cup \Psi_{f^2} \cup \Psi_{a^2} \cup \Psi_{f,a} \cup \Psi_{f,a^2} \cup \Psi_{f^2,a^2}$. 

\subsubsection{Proximity Feature Extraction with Meta Diagram}

Given a pair of users, e.g., $u_i^{(1)}$ and $u_j^{(2)}$, based on meta diagram $\Phi_k \in \Phi$, we can represent the set of meta diagram instances connecting $u_i^{(1)}$ and $u_j^{(2)}$ as $\mathcal{P}_{\Phi_k}(u_i^{(1)}, u_j^{(2)})$. Users $u_i^{(1)}$ and $u_j^{(2)}$ can have multiple meta diagram instances going into/out from them. Formally, we can represent all the meta diagram instances going out from user $u_i^{(1)}$ (or going into $u_j^{(2)}$), based on meta diagram $\Phi_k$, as set $\mathcal{P}_{\Phi_k}(u_i^{(1)}, \cdot)$ (or $\mathcal{P}_{\Phi_k}(\cdot, u_j^{(2)})$). The proximity score between $u_i^{(1)}$ and $u_j^{(2)}$ based on meta diagram $\Phi_k$ can be represented as the following \textit{meta proximity} concept formally.

\noindent \textbf{Definition 6} (Meta Diagram Proximity): Based on meta diagram $\Phi_k$, the meta diagram proximity between users $u_i^{(1)}$ and $u_j^{(2)}$ in $G$ can be represented as
\vspace{-6pt}
$$s_{\Phi_k}(u_i^{(1)}, u_j^{(2)}) = \frac{2|\mathcal{P}_{\Phi_k}(u_i^{(1)}, u_j^{(2)})|}{|\mathcal{P}_{\Phi_k}(u_i^{(1)}, \cdot)| + |\mathcal{P}_{\Phi_k}(\cdot, u_j^{(2)})|}.$$

\textit{Meta diagram proximity} considers not only the meta diagram instances between $u_i^{(1)}$ and $u_j^{(2)}$ but also penalizes those going out from and into $u_i^{(1)}$ and $u_j^{(2)}$, respectively, at the same time. Since the meta diagrams span the whole network, both the local and global network structure can be captured by the the meta diagrams. With the above meta proximity definition, we can represent the meta proximity scores among all users in the network $G$ based on meta diagram $\Phi_k$ as matrix $\mb{S}_{\Phi_k} \in \mathbb{R}^{|\mathcal{U}| \times |\mathcal{U}|}$, where entry ${S}_{\Phi_k}(i,j) = s_{\Phi_k}(u_i^{(1)}, u_j^{(2)})$. All the meta proximity matrices defined for network $G$ can be represented as $\{\mb{S}_{\Phi_k}\}_{\Phi_k \in \Phi}$. 

Meanwhile, according to the \textit{meta proximity} definition, to compute the proximity scores among users, we need to count the number of \textit{meta diagram} instances connecting users. However, different from the meta path instance counting (which can be done in polynomial time), counting the number of \textit{meta diagram} instances among users is never an easy task. It involves the graph isomorphism step to match subnetworks with the \textit{meta diagram} structure and node/link types. To lower down the computational time costs, we propose the \textit{minimum meta diagram covering set} concept, which will be applied to shrink the search space of nodes in the networks. 

\noindent \textbf{Definition 7} (Meta Diagram Covering Set): Give a \textit{anchor meta diagram} $\Psi$ starting and ending with node types $n_s$ and $n_t$, $\Psi$ will contain multiple paths connecting $n_s$ and $n_t$. Formally, these covered paths connecting $n_s$ and $n_t$ can be represented as the covering set of $\Psi$, i.e., $\mathcal{C}(\Psi) = \{\Rho_1, \Rho_2, \cdots, \Rho_n\}$, where $\Rho_i \in \mathcal{C}(\Psi)$ denotes a \textit{meta path} from $n_s$ to $n_t$. Anchor meta diagram $\Psi$ can be decomposed in different ways, and we are only interested in the \textit{minimum meta diagram covering set} with the smallest size $|\mathcal{C}(\Psi)|$. The the anchor meta diagram covering set recovers the set of meta paths composing the diagrams as introduced before, which can clearly indicate the relationship between meta path and meta diagram.

\noindent \textbf{LEMMA 1}: Given a \textit{meta diagram} $\Psi$, a pair of nodes $u_i^{(1)}, u_j^{(2)} \subset \mathcal{V}$ are connected by instances of \textit{meta diagram} $\Psi$ iff $u_i^{(1)}, u_j^{(2)}$ can be connected by instances of all meta paths in its covering set $\mathcal{C}(\Psi)$.

\noindent \textbf{PROOF}:
The lemma can be proved by contradiction. Let's assume the lemma doesn't hold, and $\exists \Rho_k \in \mathcal{C}(\Psi)$ that cannot connect $u_i^{(1)}, u_j^{(2)}$ in the network, given that $\Psi$ has an instance connecting $u_i^{(1)}, u_j^{(2)}$. Since $\Rho_k$ is one part of $\Psi$, and we can identify the corresponding parts of $\Rho_k$ from $\Psi$'s instance, which will create a path connecting $u_i^{(1)}$ with $u_j^{(2)}$. It contradicts the assumption. Therefore, the Lemma should hold.


Furthermore, based on the above Lemma 1, we can also derive the relationship between the covering sets of meta diagrams.

\noindent \textbf{LEMMA 2}: Given two \textit{meta diagrams} $\Psi_i$ and $\Psi_j$, where $\mathcal{C}(\Psi_i) \subseteq \mathcal{C}(\Psi_j)$, if a pair of nodes $u_i^{(1)}, u_j^{(2)} \subset \mathcal{V}$ can be connected by instances of \textit{meta diagram} $\Psi_j$, there will also be an instance of \textit{meta diagram} $\Psi_i$ connecting $u_i^{(1)}, u_j^{(2)}$ in the network as well.

The above lemma can be proved in a similar way as the proof of Lemma 1, which will not be introduced here due to the limited space. Based on the above lemmas, we propose to apply the \textit{meta diagram covering set} to help shrink the search space. First of all, we can compute the set of meta path instances connecting users across networks. Formally, given a \textit{meta diagram} $\Psi_k$, we can obtain its minimum covering set $\mathcal{C}(\Psi_k)$. For each meta path in $\mathcal{C}(\Psi)$, a set of meta path instances connecting the input node pairs can be extracted. By combining these meta path instances together and checking their existence in the network, we will extract instances of $\Psi$. Furthermore, in the case that there exist a prior computation result of meta diagram $\Psi_{k'}$ with covering set $\mathcal{C}(\Psi_{k'}) \subset \mathcal{C}(\Psi_k)$, instead of recompute the diagram instances based on meta paths in $\mathcal{C}(\Psi)$, we can just combine the meta diagram instances of $\Psi_{k'}$ and the instances of meta paths in $\mathcal{C}(\Psi_k) \setminus \mathcal{C}(\Psi_{k'})$ to get the meta diagram instance for $\Psi_k$.


\subsection{Active Network Alignment Model}

In this part, we will introduce the \textit{active network alignment} model {\our} for the anchor link prediction across networks, which involves $4$ main components: (1) \textit{discriminative function} for labeled instances, (2) \textit{generative function} for unlabeled instance, (3) \textit{one-to-one constraint} modeling, and (4) \textit{active query} component.

\subsubsection{Labeled Data Discriminative Loss Function}

For all the potential anchor links in set $\mathcal{H}$ (involving both the labeled and unlabeled anchor link instances), a set of features will be extracted based on the meta diagrams introduced before. Formally, the feature vector extracted for anchor link $l \in \mathcal{H}$ can be represented as vector $\mb{x}_l \in \mathbb{R}^{d}$ (parameter $d$ denotes the feature size). Meanwhile, we can denote the label of link $l \in \mathcal{L}$ as $y_l \in \mathcal{Y}$ ($\mathcal{Y} = \{0, +1\}$), which denotes the existence of anchor link $l$ between the networks. For the existing anchor links in set $\mathcal{L}_+$, they will be assigned with $+1$ label; while the labels of anchor links in $\mathcal{U}$ are unknown. All the labeled anchor links in set $\mathcal{L}_+$ can be represented as a tuple set $\{(\mb{x}_l, y_l)\}_{l \in \mathcal{L}_+ }$. Depending on whether the anchor link instances are linearly separable or not, the extracted anchor link feature vectors can be projected to different feature spaces with various kernel functions $g: \mathbb{R}^d \to \mathbb{R}^k$. For instance, given the feature vector $\mb{x}_l \in \mathbb{R}^d$ of anchor link $l$, we can represent its projected feature vector as $g(\mb{x}_l) \in \mathbb{R}^k$. In this paper, the linear kernel function will be used for simplicity, and we have $g(\mb{x}_l) = \mb{x}_l$ for all the links $l$. 

In the \textit{active network alignment} model, the \textit{discriminative} component can effectively differentiate the positive instances from the non-existing ones, which can be denoted as mapping $f(\cdot; \mb{\theta}_f): \mathbb{R}^d \to \{+1, 0\}$ parameterized by $\mb{\theta}_f$. In this paper, we will use a linear model to fit the link instances, and the \textit{discriminative} model to be learned can be represented as $f(\mb{x}_l; {\mb{w}}) = \mb{w}^\top \mb{x}_l + b$, where $\mb{\theta}_f = [\mb{w}, b]$. By adding a dummy feature $1$ for all the anchor link feature vectors, we can incorporate bias term $b$ into the weight vector $\mb{w}$ and the parameter vector can be denoted as $\mb{\theta}_f = \mb{w}$ for simplicity. Based on the above descriptions, we can represent the introduced \textit{discriminative} loss function on the labeled set $\mathcal{L}_+$ as
$$L(f, \mathcal{L}_+ ; \mb{w}) = \sum_{l \in \mathcal{L}_+ } \big( f(\mb{x}_l; \mb{w}) - y_l \big)^2 = \sum_{l \in \mathcal{L}_+ } (\mb{w}^\top \mb{x}_l - y_l)^2.$$

\subsubsection{Unlabeled Data Generative Loss Function}

Meanwhile, to alleviate the insufficiency of labeled data, we also propose to utilize the unlabeled anchor links to encourage the learned model can capture the salient structures of all the anchor link instances. Based on the above discriminative model function $f(\cdot; \mb{w})$, for a unlabeled anchor link $l \in \mathcal{U}$, we can represent its inferred ``label'' as $y_l = f(\mb{x}_l; \mb{w})$. Considering that the result of $f(\cdot; \mb{w})$ may not necessary the exact label values in $\mathcal{Y}$, in the \textit{generative} component, we can represent the generated anchor link label as $sign\big(f(\mb{x}_l; \mb{w})\big) \in \{+1, 0\}$. How to determine its value will be introduced later in Section~\ref{subsec:joint}. Based on it, the loss function introduced in the \textit{generative} component based on the unlabeled anchor links can be denoted as \vspace{-4pt}
$$L(f, \mathcal{U}; \mb{w}) = \sum_{l \in \mathcal{U}} \Big( \mb{w}^\top \mb{x}_l - sign\big(f(\mb{x}_l; \mb{w}) \big) \Big)^2.$$

\subsubsection{Query Component and Query Loss Function}

Furthermore, besides the labeled links, a subset of the anchor links in $\mathcal{U}$ will be selected to query for the labels from the oracle, which can be denoted as set $\mathcal{U}_q$ formally. The true label of anchor link $l \in \mathcal{U}_q$ after query can be represented as $\tilde{y}_l \in \{+1, 0\}$. The remaining anchor links in set $\mathcal{U}$ can be represented as $\mathcal{U} \setminus \mathcal{U}_q$. Based on the loss functions introduced before, depending on whether the labels of links are queried or not, we can further specify the loss function for set $\mathcal{U}$ as 
\begin{align*}
&L(f, \mathcal{U}; \mb{w}) = L(f, \mathcal{U}_q; \mb{w}) + L(f, \mathcal{U} \setminus \mathcal{U}_q; \mb{w})\\
&= \sum_{l \in \mathcal{U}_q} (\mb{w}^\top \mb{x}_l - \tilde{y}_l)^2 +  \sum_{l \in \mathcal{U} \setminus \mathcal{U}_q} \Big(\mb{w}^\top \mb{x}_l - sign\big(f(\mb{x}_l; \mb{w}) \big)\Big)^2.
\end{align*}
Here, we need to add more remarks that notation $\tilde{y}_l$ denotes the queried label of anchor link $l \in \mathcal{U}_q$ which will be a known value, while the labels for the remaining anchor link $l \in \mathcal{U}\setminus \mathcal{U}_q$ will to be inferred in the model.

\subsubsection{Cardinality Mathematical Constraint}

As introduced before, the anchor links to be inferred between networks are subject to the \textit{one-to-one} cardinality constraint. Such a constraint will control the maximum number of anchor links incident to the user nodes in each networks. Subject to the cardinality constraints, the prediction task of anchor links between networks are no longer independent. For instance, if anchor link $(u^{(1)}, v^{(2)})$ is predicted to be positive, then all the remaining anchor links incident to $u^{(1)}$ and $v^{(2)}$ in the unlabeled set $\mathcal{U}$ will be negative by default. Viewed in such a perspective, the cardinality constraint on anchor links should be effectively incorporated in model building, which will be modeled as the mathematical constraints on node degrees in this paper. To represent the user node-anchor link relationships in networks $G^{(1)}$ and $G^{(2)}$ respectively, we introduce the user node-anchor link incidence matrices $\mb{A}^{(1)}  \in \{0, 1\}^{|\mathcal{U}^{(1)}| \times |\mathcal{H}|}, \mb{A}{(2)} \in \{0, 1\}^{|\mathcal{U}^{(2)}| \times |\mathcal{H}|}$ in this paper. Entry $A^{(1)}(i, j) = 1$ iff anchor link $l_j \in \mathcal{H}$ is connected with user node $u_i^{(1)}$ in $G^{(1)}$, and it is similar for $A^{(2)}$.

According to the analysis provided before, we can represent the labels of links in $\mathcal{H}$ as vector $\mb{y} \in \{+1, 0\}^{|\mathcal{H}|}$, where entry $y(i)$ represents the label of link $l_i \in \mathcal{L}$. Depending on which group $l_i$ belongs to, its value has different representations as introduced before $y(i) = +1, \mbox{ if } l_i \in \mathcal{L}_+$; $y(i) \tilde{y}_{l_i}, \mbox{ if } l_i \in \mathcal{U}_q$, and $y(i)$ is unknown if $l_i \in \mathcal{U} \setminus \mathcal{U}_q$. Furthermore, based on the anchor link label vector $\mb{y}$, user node-anchor link incidence matrices $\mb{A}^{(1)}$ and $\mb{A}^{(2)}$, we can represent the user node degrees in networks $G^{(1)}$ and $G^{(2)}$ as vectors $\mb{d}^{(1)} \in \mathbb{N}^{|\mathcal{H}|}$ and $\mb{d}^{(2)} \in \mathbb{N}^{|\mathcal{H}|}$ respectively.
\begin{alignat*}{2}
\mb{d}^{(1)} = \mb{A}^{(1)} \mb{y}
\mbox{, and }
\mb{d}^{(2)} = \mb{A}^{(2)} \mb{y}.
\end{alignat*}

Therefore, the \textit{one-to-one} constraint on anchor links can be denoted as the constraints on node degrees in $G^{(1)}$ and $G^{(2)}$ as follows:
\begin{alignat*}{2}
\mb{0} \le \mb{A}^{(1)} \mb{y} \le \mb{1}
\mbox{, and }
\mb{0} \le \mb{A}^{(2)} \mb{y} \le \mb{1}.
\end{alignat*}

\subsection{Joint Optimization Objective Function}\label{subsec:joint}

Based on the introduction in the previous subsection, by combining the loss terms introduced by the labeled, queried and remaining unlabeled anchor links together with the cardinality constraint, we can represent the joint optimization objective function as
\begin{align*}
\min_{\mb{w}, \mb{y}, \mathcal{U}_q} &L(f, \mathcal{L}_+ ; \mb{w}) + \alpha \cdot L(f, \mathcal{U}_q; \mb{w}) \\
&\ \ \ \ \ \ \ \ \ \ \ \ \ \ \ + \beta \cdot L(f, \mathcal{U} \setminus \mathcal{U}_q; \mb{w}) + \gamma \cdot \left\|\mb{w}\right\|_2^2\\
&s.t.\ \  |\mathcal{U}_q| \le b \mbox{, and } y_l = \tilde{y}_l, \forall l \in \mathcal{U}_q, \\
&\ \ \ \ \ \ y_l \in \{+1, 0\}, \forall l \in \mathcal{U} \setminus \mathcal{U}_q,\mbox{ and } y_l = + 1, \forall l \in \mathcal{L}_+,\\
&\ \ \ \ \ \  \mb{0} \le \mb{A}^{(1)} \mb{y} \le \mb{1} \mbox{, and } \mb{0} \le \mb{A}^{(2)} \mb{y} \le \mb{1}.
\end{align*}

Here, we set the weight scalar $\alpha$ and $\beta$ with the value $1$, because we assume that each link is equally important for training, if no other external knowledge exists, regardless of whether it belongs to $\mathcal{U}_q$ or $\mathcal{U}\setminus \mathcal{U}_q$. In this way, the new loss term of all the links in sets $\mathcal{L}_+$, $\mathcal{U}_q$ and $\mathcal{U} \setminus \mathcal{U}_q$ can be simplified as

\begin{align*}
&L(f, \mathcal{L}_+ ; \mb{w}) + \alpha \cdot L(f, \mathcal{U}_q; \mb{w}) + \beta \cdot L(f, \mathcal{U} \setminus \mathcal{U}_q; \mb{w}) \\
&= L(f, \mathcal{H}; \mb{w}) =  \left\| \mb{X} \mb{w} - \mb{y} \right\|_2^2,
\end{align*}
where matrix $\mb{X} = [\mb{x}_{l_1}^\top, \mb{x}_{l_2}^\top, \cdots, \mb{x}_{l_{|\mathcal{H}|}}^\top]^T$ denotes the feature matrix of all the links in set $\mathcal{H}$.

Here, we can see the objective function involve multiple variables, i.e., variable $\mb{w}$, label $\mb{y}$, and the query set $\mathcal{U}_q$, and the objective is not jointly convex with regarding these variables. What's more, the inference of the label variable $\mb{y}$ and the query set $\mathcal{U}_q$ are both combinatorial problems, and obtaining their optimal solution will be NP-hard. In this paper, we design an hierarchical alternative variable updating process for solving the problem instead:
\begin{enumerate}
	\item fix $\mathcal{U}_q$, and update $\mb{y}$ and $\mb{w}$,
	\begin{enumerate}[label={(1-\arabic*)}]
		\item with fixed $\mathcal{U}_q$, fix $\mb{y}$, update $\mb{w}$,
		\item with fixed $\mathcal{U}_q$, fix $\mb{w}$, update $\mb{y}$,
	\end{enumerate}
	\item fix $\mb{y}$ and $\mb{w}$, and update $\mathcal{U}_q$. 
\end{enumerate}
A remark to be added here: we can see that variable $\mathcal{U}_q$ is different from the remaining two, which involves the label query process with the oracle subject to the specified budget. To differentiate these two iterations, we call the iterations (1) and (2) as the \textit{external iteration}, while call (1-1) and (1-2) \textit{internal iteration}. Next, we will illustrate the detailed alternative learning algorithm as follows.

\noindent $\bullet$ \textbf{External Iteration Step (1)}: Fix $\mathcal{U}_q$, Update $\mb{y}$, $\mb{w}$.

{\tiny $\blacksquare$} \textbf{Internal Iteration Step (1-1)}: Fix $\mathcal{U}_q$, $\mb{y}$, Update $\mb{w}$.

With $\mb{y}$, $\mathcal{U}_q$ fixed, we can represent the objective function involving variable $\mb{w}$ as
\begin{align*}
\min_{\mb{w}} \frac{c}{2} \left\| \mb{X} \mb{w} - \mb{y} \right\|_2^2 + \frac{1}{2} \left\|\mb{w}\right\|_2^2.
\end{align*}

The objective function is a quadratic convex function, and its optimal solution can be represented as
$$ \mb{w} = \mb{H} \mb{y} = c(\mb{I} + c\mb{X}^\top \mb{X})^{-1} \mb{X}^\top \mb{y},$$
where $\mb{H} = c(\mb{I} + c\mb{X}^\top \mb{X})^{-1} \mb{X}^\top$ is a constant matrix. Therefore, the weight vector $\mb{w}$ depends only on the $\mb{y}$ variable.

{\tiny $\blacksquare$} \textbf{Internal Iteration Step (1-2)}: Fix $\mathcal{U}_q$, $\mb{w}$, Update $\mb{y}$.

With $\mathcal{U}_q$, $\mb{w}$ fixed, together with the constraint, we know that terms $L(f, \mathcal{L}_+ ; \mb{w})$, $L(f, \mathcal{U}_q; \mb{w})$ and $\left\| \mb{w} \right\|_2^2$ are all constant. And the objective function will be
\begin{align*}
&\min_{\mb{y}} \left\| \mb{X}\mb{w} - \mb{y} \right\|_2^2 \\
&s.t. \ y_l \in \{+1, 0\}, \forall l \in \mathcal{U} \setminus \mathcal{U}_q,\\
&\ \ \ \ \ y_l = \tilde{y}_l, \forall l \in \mathcal{U}_q \ \mbox{and} \ y_l = + 1, \forall l \in \mathcal{L}_+ \mbox{,} \\
&\ \ \ \ \ \mb{0} \le \mb{A}^{(1)} \mb{y} \le \mb{1} \mbox{, and } \mb{0} \le \mb{A}^{(2)} \mb{y} \le \mb{1}.
\end{align*}

It is an integer programming problem, which has been shown to be NP-hard and no efficiently algorithm exists that lead to the optimal solution. In this paper, we will use the greedy link selection algorithm proposed in \cite{ZCZCY17} based on values $\hat{\mb{y}} =  \mb{X}\mb{w}$, which has been proven to achieve $\frac{1}{2}$-approximation of the optimal solution. The time complexity of this step is $O(|\tilde{L}|)$, where $\tilde{L} = \{l|l\in\mathcal{U}\setminus \mathcal{U}_q\}$.

\noindent $\bullet$ \textbf{External Iteration Step (2)}: Fix $\mb{w}$, $\mb{y}$, Update $\mathcal{U}_q$.


Selecting the optimal set $\mathcal{U}_q$ at one time involves the search of all the potential $b$ link instance combinations from the unlabeled set $\mathcal{U}$, whose search space is $\dbinom{|\mathcal{U}|}{b}$, and there is no known efficient approach for solving the problem in polynomial time. Therefore, instead of selecting them all at one time, we propose to choose several link instances greedily in each iterations. Due to the one-to-one constraint, the unlabeled anchor links no longer bears equal information, and querying for labels of potential positive anchor links will be more ``informative'' compared with negative anchor links. Among the unlabeled links, {\our} selects a set of mis-classified false-negative anchor links (but with a large positive score) as the potential candidates, benefits introduced by whose label queries includes both their own label corrections and other extra label gains of their conflicting negative links at the same time. Formally, among all the unlabeled links in $\mathcal{U}$, we can represent the set of links classified to be positive/negative instances in the previous iteration step as $\mathcal{U}^+ = \{l | l \in \mathcal{U}, y_l = +1\}$ and $\mathcal{U}^- = \{l | l \in \mathcal{U}, y_l = 0\}$. Based on these two sets, the group of potentially mis-classified false-negative anchor link candidates as set 
$$\mathcal{C} = \{l | l \in \mathcal{U}^-, \exists l', l'' \in \mathcal{U}^+ \mbox{ that conflicts with } l, $$
$$\ \ \ \ \ \ \ \ \ \ \ \ \ \ \ \ \ \ \ \ \ \ \ \ \ \ \ \ \ \ \ \ \ \hat{y}_{l'} \sim \hat{y_l} \gg  \hat{y}_{l''} > 0\},$$
where statement ``$l'$/$l''$ conflicts with $l$'' denotes $l'$/$l''$ and $l$ are incident to the same nodes respectively. Operator $\hat{y}_{l'} \sim \hat{y_l}$ represents $\hat{y}_{l'}$ is close to $\hat{y_l}$ (whose difference threshold is set as $0.05$ in the experiments). All the links in set $\mathcal{C}$ will be sorted according to value $\hat{y_l}-\hat{y}_{l''}$, and, instead of adding one by one, the top $k$ candidates will be added to $\mathcal{U}_q$ in this iteration (Here, $k$ denotes the query batch size, which is assigned with value $5$ in the experiments). Because {\our} has to select the top $k$ candidates from all potential candidates, where the potential candidates we defined as $\tilde{L}^- = \{l|l\in\mathcal{U}\setminus \mathcal{U}_q, \tilde{y_l} = 0\}$, the time complexity of External Iteration Step (2) is $O(|\tilde{L}^-|)$.

\subsection{Time Complexity Analysis}
Here, we start to analyze the time complexity of {\our} from a holistic perspective based on the analysis of each step in section \ref{subsec:joint}. As we set the query batch size as $k$ and the budget as $b$, the whole hierarchical alternative variable updating process has to be executed $b/k$ rounds. The iteration step (1-1) is a matrix multiplication which has he time complexity $O(d*|\mathcal{H}|)$. The time complexity the iteration step (1-2) is $O(|\tilde{L}|)$. Besides, the time complexity of the iteration step (2) is $O(|\tilde{L}^-|)$.We can find {\our} is scalable, with near linear runtime in the data size $|\mathcal{H}|$.

\section{Experiments}\label{sec:experiment}

To demonstrate the effectiveness of {\our} and the meta diagram based features, extensive experiments have been done on real-world  heterogeneous social networks. In the following part, we will describe the dataset we use in experiments at first. Then we will introduce the experimental settings, including different comparison methods and evaluation metrics used in the experiments. At last, we will show the experimental results together with the convergence analysis and parameter sensitivity analysis.  

\subsection{Dataset Description}
\begin{table}[t]
	\caption{Properties of the Heterogeneous Networks}
	\label{tab:datastat}
	\centering
	\small
	\begin{tabular}{clrr}
		\toprule
		&&\multicolumn{2}{c}{network}\\
		\cmidrule{3-4}
		&property &\textbf{Twitter} &\textbf{Foursquare}   \\
		\midrule 
		\multirow{3}{*}{\# node}
		&user   & 5,223 & 5,392 \\
		&tweet/tip  & 9,490,707 & 48,756 \\
		&location & 297,182 & 38,921 \\
		\midrule 
		\multirow{2}{*}{\# link}
		&friend/follow    &164,920  &76,972 \\
		&write    & 9,490,707 & 48,756 \\
		\bottomrule
	\end{tabular}
	\vspace{-10pt}
\end{table}

Our dataset used in experiments consists of two heterogeneous networks: Foursquare and Twitter. Both of them are famous online social networks. The key statistical data describing these two networks can be found in Table~\ref{tab:datastat}. About the method and strategy of crawling this dataset, you can get detailed information in \cite{KZY13, ZKY13}.

\begin{table*}[t]
	\caption{Performance comparison of different methods for Network Alignment. We use different NP-ratios with $\gamma=60\%$.}
	\label{tab:main_result_fix_train_ratio}
	\centering
	\setlength{\tabcolsep}{2.5pt}
	{\scriptsize
		\begin{tabular}{lrcccccccccc}
			\toprule
			\multicolumn{2}{l}{}&\multicolumn{8}{c}{Negative Positive Ratio $\theta$}\\
			\cmidrule{3-12}
			metrics &methods	&5	& 10	& 15	& 20	&25	&30	&35	& 40	& 45 & 50 \\
			\midrule
			\multirow{6}{*}{\rotatebox{90}{F1}}
			&{\ourtwentyround}	&\textbf{0.631}$\pm$\textbf{0.01}     &\textbf{0.575}$\pm$\textbf{0.01}     &\textbf{0.524}$\pm$\textbf{0.01}     &\textbf{0.484}$\pm$\textbf{0.02}     &\textbf{0.455}$\pm$\textbf{0.02}     &\textbf{0.436}$\pm$\textbf{0.02}     &\textbf{0.413}$\pm$\textbf{0.01}     &\textbf{0.402}$\pm$\textbf{0.02}     &\textbf{0.384}$\pm$\textbf{0.01}     &\textbf{0.363}$\pm$\textbf{0.01}	\\
			&{\ourtenround}&0.625$\pm$0.01       &0.571$\pm$0.01       &0.514$\pm$0.01       &0.482$\pm$0.02       &0.454$\pm$0.02       &0.429$\pm$0.02       &0.404$\pm$0.01       &0.392$\pm$0.02       &0.374$\pm$0.02       &0.361$\pm$0.01 	\\
			\cmidrule{3-12}
			&{\activerandom}	&0.616$\pm$0.01       &0.553$\pm$0.01       &0.501$\pm$0.01       &0.463$\pm$0.01       &0.437$\pm$0.01       &0.413$\pm$0.01       &0.392$\pm$0.02       &0.381$\pm$0.02       &0.368$\pm$0.02       &0.352$\pm$0.01	\\
			\cmidrule{3-12}
			&{\pusvm}	&0.616$\pm$0.01       &0.556$\pm$0.01       &0.507$\pm$0.01       &0.469$\pm$0.02       &0.441$\pm$0.01       &0.414$\pm$0.02       &0.396$\pm$0.01       &0.380$\pm$0.03       &0.365$\pm$0.01       &0.350$\pm$0.01	\\
			\cmidrule{3-12}
			&{\svmmpmd}	&0.387$\pm$0.05       &0.300$\pm$0.04       &0.247$\pm$0.04       &0.165$\pm$0.06       &0.159$\pm$0.06       &0.150$\pm$0.03       &0.152$\pm$0.04       &0.102$\pm$0.06       &0.091$\pm$0.07       &0.049$\pm$0.06	\\
			&{\svmmp}	&0.476$\pm$0.11       &0.093$\pm$0.08       &0.055$\pm$0.05       &0.004$\pm$0.01       &0.002$\pm$0.00       &0.000$\pm$0.00       &0.000$\pm$0.00       &0.000$\pm$0.00       &0.000$\pm$0.00       &0.000$\pm$0.00	\\
			\midrule
			\multirow{6}{*}{\rotatebox{90}{Precision}}
			&{\ourtwentyround}	&\textbf{0.856}$\pm$\textbf{0.01}     &\textbf{0.767}$\pm$\textbf{0.01}     &\textbf{0.693}$\pm$\textbf{0.01}     &\textbf{0.632}$\pm$\textbf{0.02}     &\textbf{0.591}$\pm$\textbf{0.02}     &\textbf{0.559}$\pm$\textbf{0.02}     &\textbf{0.526}$\pm$\textbf{0.02}     &\textbf{0.509}$\pm$\textbf{0.02}     &\textbf{0.486}$\pm$\textbf{0.02}     &\textbf{0.457}$\pm$\textbf{0.02}	\\
			&{\ourtenround}	&0.848$\pm$0.01       &0.762$\pm$0.01       &0.676$\pm$0.02       &0.626$\pm$0.02       &0.587$\pm$0.02       &0.551$\pm$0.02       &0.515$\pm$0.02       &0.496$\pm$0.03       &0.473$\pm$0.02       &0.454$\pm$0.02		\\
			\cmidrule{3-12}
			&{\activerandom}	&0.836$\pm$0.01       &0.735$\pm$0.01       &0.657$\pm$0.01       &0.600$\pm$0.02       &0.563$\pm$0.02       &0.528$\pm$0.02       &0.498$\pm$0.03       &0.481$\pm$0.02       &0.462$\pm$0.02       &0.440$\pm$0.02	\\
			\cmidrule{3-12}
			&{\pusvm}		&0.835$\pm$0.01       &0.738$\pm$0.01       &0.665$\pm$0.01       &0.609$\pm$0.02       &0.569$\pm$0.02       &0.530$\pm$0.02       &0.504$\pm$0.02       &0.4809$\pm$0.02       &0.459$\pm$0.02       &0.439$\pm$0.02       \\
			\cmidrule{3-12}
			&{\svmmpmd}	&0.743$\pm$0.06       &0.703$\pm$0.04       &0.652$\pm$0.06       &0.587$\pm$0.20       &0.585$\pm$0.09       &0.520$\pm$0.05       &0.519$\pm$0.06       &0.487$\pm$0.25       &0.331$\pm$0.27       &0.311$\pm$0.31       \\
			&{\svmmp}	&0.571$\pm$0.02       &0.338$\pm$0.28       &0.323$\pm$0.27       &0.057$\pm$0.17       &0.018$\pm$0.05       &0.000$\pm$0.00       &0.000$\pm$0.00       &0.000$\pm$0.00       &0.000$\pm$0.00       &0.000$\pm$0.00\\
			\midrule
			\multirow{6}{*}{\rotatebox{90}{Recall}}
			&{\ourtwentyround}	&\textbf{0.499}$\pm$\textbf{0.01}     &\textbf{0.460}$\pm$\textbf{0.01}     &\textbf{0.422}$\pm$\textbf{0.01}     &\textbf{0.392}$\pm$\textbf{0.01}     &\textbf{0.371}$\pm$\textbf{0.01}     &\textbf{0.357}$\pm$\textbf{0.01}     &\textbf{0.339}$\pm$\textbf{0.01}     &\textbf{0.332}$\pm$\textbf{0.01}     &\textbf{0.318}$\pm$\textbf{0.01}     &\textbf{0.301}$\pm$\textbf{0.01}	\\
			&{\ourtenround}	&0.495$\pm$0.01       &0.457$\pm$0.01       &0.414$\pm$0.01       &0.392$\pm$0.01       &0.371$\pm$0.01       &0.352$\pm$0.02       &0.333$\pm$0.01       &0.324$\pm$0.01       &0.310$\pm$0.01       &0.300$\pm$0.01      \\
			\cmidrule{3-12}
			&{\activerandom}&0.488$\pm$0.01       &0.443$\pm$0.01       &0.404$\pm$0.01       &0.376$\pm$0.01       &0.357$\pm$0.01       &0.340$\pm$0.01       &0.323$\pm$0.01       &0.315$\pm$0.01      &0.305$\pm$0.01       &0.293$\pm$0.01	 \\
			\cmidrule{3-12}
			&{\pusvm}	&0.488$\pm$0.01       &0.446$\pm$0.01       &0.410$\pm$0.01       &0.381$\pm$0.02       &0.360$\pm$0.01       &0.340$\pm$0.01       &0.327$\pm$0.01       &0.314$\pm$0.01       &0.302$\pm$0.01       &0.290$\pm$0.01\\
			\cmidrule{3-12}
			&{\svmmpmd}	&0.271$\pm$0.07       &0.194$\pm$0.04       &0.155$\pm$0.03       &0.097$\pm$0.03       &0.094$\pm$0.03       &0.086$\pm$0.02       &0.088$\pm$0.02       &0.059$\pm$0.04       &0.053$\pm$0.04       &0.027$\pm$0.03\\
			&{\svmmp}	&0.439$\pm$0.14       &0.055$\pm$0.05       &0.031$\pm$0.03       &0.002$\pm$0.00       &0.001$\pm$0.01       &0.000$\pm$0.00       &0.000$\pm$0.00       &0.000$\pm$0.00       &0.000$\pm$0.00       &0.000$\pm$0.00\\
			\midrule
			\multirow{6}{*}{\rotatebox{90}{Accuracy}}
			&{\ourtwentyround}	&\textbf{0.902}$\pm$\textbf{0.00}     &\textbf{0.938}$\pm$\textbf{0.00}     &\textbf{0.952}$\pm$\textbf{0.01}     &\textbf{0.960}$\pm$\textbf{0.00}     &\textbf{0.966}$\pm$\textbf{0.00}     &\textbf{0.970}$\pm$\textbf{0.00}     &\textbf{0.973}$\pm$\textbf{0.00}     &\textbf{0.976}$\pm$\textbf{0.00}     &0.978$\pm$0.00     &0.979$\pm$0.00	\\
			&{\ourtenround}	&0.901$\pm$0.00       &0.938$\pm$0.00       &0.951$\pm$0.00       &0.960$\pm$0.00       &0.966$\pm$0.00       &0.970$\pm$0.00       &0.972$\pm$0.00       &0.975$\pm$0.00       &0.977$\pm$0.00       &0.979$\pm$0.00       \\
			\cmidrule{3-12}
			&{\activerandom}	&0.898$\pm$0.00       &0.934$\pm$0.00       &0.949$\pm$0.00       &0.958$\pm$0.00       &0.964$\pm$0.00       &0.968$\pm$0.00       &0.972$\pm$0.00       &0.975$\pm$0.00       &0.977$\pm$0.00       &0.978$\pm$0.00\\
			\cmidrule{3-12}
			&{\pusvm}	&0.898$\pm$0.00       &0.935$\pm$0.00       &0.950$\pm$0.00       &0.958$\pm$0.00       &0.964$\pm$0.00       &0.969$\pm$0.00       &0.972$\pm$0.00      &0.975$\pm$0.00       &0.977$\pm$0.00       &0.978$\pm$0.00\\
			\cmidrule{3-12}
			&{\svmmpmd}	&0.860$\pm$0.00       &0.918$\pm$0.00       &0.941$\pm$0.00       &0.954$\pm$0.00       &0.962$\pm$0.00       &0.968$\pm$0.00       &0.972$\pm$0.00       &0.976$\pm$0.00       &\textbf{0.978}$\pm$\textbf{0.00}       &\textbf{0.980}$\pm$\textbf{0.00}       \\
			&{\svmmp}	&0.850$\pm$0.00       &0.909$\pm$0.00       &0.937$\pm$0.00       &0.952$\pm$0.00       &0.961$\pm$0.00      &0.967$\pm$0.00      &0.972$\pm$0.00       &0.975$\pm$0.00       &0.978$\pm$0.00      &0.980$\pm$0.00	\\
			
			\bottomrule
		\end{tabular}\vspace{-14pt}
	}
	
\end{table*}

\begin{table*}[t]
	\caption{Performance comparison of different methods for Network Alignment. We use different sample-ratios with $\theta = 50$.} 
	\label{tab:main_result_fix_np_ratio}
	\centering
	\setlength{\tabcolsep}{2.5pt}
	{\scriptsize
		\begin{tabular}{lrcccccccccc}
			\toprule
			\multicolumn{2}{l}{}&\multicolumn{8}{c}{Sample Ratio $\gamma$}\\
			\cmidrule{3-12}
			metrics &methods	&10\%&20\%	& 30\%	& 40\%	&50\%	&60\%	&70\%	& 80\%	& 90\% & 100\% \\
			\midrule
			\multirow{6}{*}{\rotatebox{90}{F1}}
			&{\ourtwentyround}	&\textbf{0.235}$\pm$\textbf{0.00}     &\textbf{0.265}$\pm$\textbf{0.02}     &\textbf{0.291}$\pm$\textbf{0.02}     &\textbf{0.309}$\pm$\textbf{0.01}     &\textbf{0.333}$\pm$\textbf{0.01}     &\textbf{0.363}$\pm$\textbf{0.01}     &\textbf{0.369}$\pm$\textbf{0.02}     &\textbf{0.397}$\pm$\textbf{0.01}     &\textbf{0.404}$\pm$\textbf{0.00}     &\textbf{0.422}$\pm$\textbf{0.01} 	\\
			&{\ourtenround}&0.230$\pm$0.01       &0.247$\pm$0.01       &0.289$\pm$0.02       &0.300$\pm$0.01       &0.323$\pm$0.02       &0.361$\pm$0.01       &0.362$\pm$0.02       &0.396$\pm$0.01       &0.399$\pm$0.00       &0.410$\pm$0.01       \\
			\cmidrule{3-12}
			&{\activerandom}	&0.219$\pm$0.01       &0.234$\pm$0.01       &0.284$\pm$0.02      &0.289$\pm$0.01       &0.316$\pm$0.01       &0.352$\pm$0.01       &0.360$\pm$0.01       &0.383$\pm$0.01       &0.391$\pm$0.00       &0.402$\pm$0.01      \\
			\cmidrule{3-12}
			&{\pusvm}&0.217$\pm$0.01       &0.233$\pm$0.01       &0.280$\pm$0.02       &0.293$\pm$0.01       &0.316$\pm$0.02       &0.350$\pm$0.01       &0.361$\pm$0.02       &0.385$\pm$0.01       &0.387$\pm$0.00       &0.400$\pm$0.01	\\
			\cmidrule{3-12}
			&{\svmmpmd}	&0.005$\pm$0.01       &0.006$\pm$0.01       &0.065$\pm$0.04       &0.043$\pm$0.05       &0.042$\pm$0.06       &0.049$\pm$0.06       &0.082$\pm$0.06       &0.09$\pm$0.06       &0.092$\pm$0.07       &0.131$\pm$0.06	\\
			&{\svmmp}	&0.005$\pm$0.01       &0.000$\pm$0.00       &0.000$\pm$0.00       &0.000$\pm$0.00       &0.000$\pm$0.00       &0.000$\pm$0.00       &0.000$\pm$0.00       &0.000$\pm$0.00       &0.000$\pm$0.00       &0.000$\pm$0.00	\\
			\midrule
			\multirow{6}{*}{\rotatebox{90}{Precision}}
			&{\ourtwentyround}	&\textbf{0.318}$\pm$\textbf{0.01}     &\textbf{0.352}$\pm$\textbf{0.02}     &0.379$\pm$0.02     &\textbf{0.396}$\pm$\textbf{0.01}     &\textbf{0.424}$\pm$\textbf{0.02}     &\textbf{0.457}$\pm$\textbf{0.02}     &\textbf{0.460}$\pm$\textbf{0.03}     &\textbf{0.491}$\pm$\textbf{0.01}     &\textbf{0.499}$\pm$\textbf{0.01}     &\textbf{0.518}$\pm$\textbf{0.02}	\\
			&{\ourtenround}&0.310$\pm$0.01       &0.327$\pm$0.02       &0.375$\pm$0.02       &0.384$\pm$0.015       &0.410$\pm$0.02       &0.45$\pm$0.02       &0.450$\pm$0.03       &0.489$\pm$0.02       &0.492$\pm$0.01       &0.503$\pm$0.02	\\
			\cmidrule{3-12}
			&{\activerandom}	&0.295$\pm$0.01       &0.310$\pm$0.01       &0.369$\pm$0.02       &0.370$\pm$0.01       &0.400$\pm$0.02       &0.440$\pm$0.02       &0.447$\pm$0.02       &0.471$\pm$0.02       &0.480$\pm$0.01       &0.493$\pm$0.01    	\\
			\cmidrule{3-12}
			&{\pusvm}		&0.292$\pm$0.01       &0.308$\pm$0.01       &0.364$\pm$0.02       &0.374$\pm$0.01       &0.399$\pm$0.02       &0.439$\pm$0.02       &0.448$\pm$0.02       &0.474$\pm$0.01       &0.475$\pm$0.01       &0.489$\pm$0.01       \\
			\cmidrule{3-12}
			&{\svmmpmd}	&0.050$\pm$0.15       &0.078$\pm$0.19       &\textbf{0.395}$\pm$\textbf{0.27}       &0.236$\pm$0.29       &0.180$\pm$0.27       &0.311$\pm$0.31       &0.343$\pm$0.28       &0.424$\pm$0.27       &0.361$\pm$0.29       &0.449$\pm$0.22 \\
			&{\svmmp}	&0.044$\pm$0.13       &0.000$\pm$0.00       &0.000$\pm$0.00       &0.000$\pm$0.00       &0.000$\pm$0.00       &0.000$\pm$0.00       &0.000$\pm$0.00       &0.000$\pm$0.00       &0.000$\pm$0.00       &0.000$\pm$0.00\\
			\midrule
			\multirow{6}{*}{\rotatebox{90}{Recall}}
			&{\ourtwentyround}	&\textbf{0.186}$\pm$\textbf{0.01}     &\textbf{0.213}$\pm$\textbf{0.01}     &\textbf{0.236}$\pm$\textbf{0.01}     &\textbf{0.253}$\pm$\textbf{0.01}     &\textbf{0.274}$\pm$\textbf{0.01}     &\textbf{0.301}$\pm$\textbf{0.01}     &\textbf{0.308}$\pm$\textbf{0.02}     &\textbf{0.334}$\pm$\textbf{0.01}     &\textbf{0.339}$\pm$\textbf{0.00}     &\textbf{0.356}$\pm$\textbf{0.01}\\
			
			&{\ourtenround}&0.183$\pm$0.01       &0.198$\pm$0.01       &0.235$\pm$0.01       &0.246$\pm$0.01       &0.267$\pm$0.01       &0.300$\pm$0.01       &0.303$\pm$0.02       &0.333$\pm$0.01       &0.336$\pm$0.01       &0.347$\pm$0.01	 \\
			\cmidrule{3-12}
			&{\activerandom}&0.174$\pm$0.01       &0.188$\pm$0.01       &0.231$\pm$0.01       &0.237$\pm$0.01       &0.261$\pm$0.01       &0.293$\pm$0.01       &0.302$\pm$0.01       &0.322$\pm$0.01       &0.330$\pm$0.00       &0.340$\pm$0.01	 \\
			\cmidrule{3-12}
			&{\pusvm}&0.173$\pm$0.01       &0.188$\pm$0.01       &0.228$\pm$0.01       &0.241$\pm$0.01       &0.261$\pm$0.01       &0.290$\pm$0.01       &0.302$\pm$0.01       &0.324$\pm$0.01       &0.327$\pm$0.00       &0.338$\pm$0.00\\
			\cmidrule{3-12}
			&{\svmmpmd}	&0.002$\pm$0.01       &0.003$\pm$0.01       &0.036$\pm$0.02       &0.024$\pm$0.03       &0.024$\pm$0.03       &0.027$\pm$0.03       &0.047$\pm$0.03       &0.056$\pm$0.03       &0.053$\pm$0.04       &0.077$\pm$0.04	\\
			&{\svmmp}	&0.003$\pm$0.01       &0.000$\pm$0.00       &0.000$\pm$0.00       &0.000$\pm$0.00       &0.000$\pm$0.00       &0.000$\pm$0.00       &0.000$\pm$0.00       &0.000$\pm$0.00       &0.000$\pm$0.00       &0.000$\pm$0.00	\\
			\midrule
			\multirow{6}{*}{\rotatebox{90}{Accuracy}}
			&{\ourtwentyround}	&0.976$\pm$0.00       &0.977$\pm$0.00       &0.977$\pm$0.00       &0.978$\pm$0.00       &0.978$\pm$0.00       &0.979$\pm$0.00       &0.979$\pm$0.00       &0.980$\pm$0.00       &0.980$\pm$0.00       &\textbf{0.981}$\pm$\textbf{0.00}  	\\
			&{\ourtenround}	&0.976$\pm$0.00       &0.976$\pm$0.00       &0.977$\pm$0.00       &0.977$\pm$0.00       &0.978$\pm$0.00       &0.979$\pm$0.00       &0.979$\pm$0.00       &0.980$\pm$0.00       &0.980$\pm$0.00       &0.980$\pm$0.00       \\
			\cmidrule{3-12}
			&{\activerandom}	&0.975$\pm$0.00       &0.975$\pm$0.00       &0.977$\pm$0.00       &0.977$\pm$0.00       &0.977$\pm$0.00       &0.978$\pm$0.00       &0.979$\pm$0.00       &0.979$\pm$0.00       &0.979$\pm$0.00       &0.980$\pm$0.00       \\
			\cmidrule{3-12}
			&{\pusvm}&0.975$\pm$0.00       &0.975$\pm$0.00       &0.977$\pm$0.00       &0.977$\pm$0.00       &0.977$\pm$0.00       &0.978$\pm$0.00       &0.979$\pm$0.00       &0.979$\pm$0.00       &0.979$\pm$0.00       &0.980$\pm$0.00       \\
			\cmidrule{3-12}
			&{\svmmpmd}	&0.980$\pm$0.00       &0.980$\pm$0.00       &0.980$\pm$0.00       &0.980$\pm$0.00       &0.980$\pm$0.00       &0.980$\pm$0.00       &0.980$\pm$0.00       &0.980$\pm$0.00       &0.980$\pm$0.00       &0.980$\pm$0.00       \\
			&{\svmmp}	&\textbf{0.980}$\pm$\textbf{0.00}     &\textbf{0.980}$\pm$\textbf{0.00}     &\textbf{0.980}$\pm$\textbf{0.00}     &\textbf{0.980}$\pm$\textbf{0.00}     &\textbf{0.980}$\pm$\textbf{0.00}     &\textbf{0.980}$\pm$\textbf{0.00}     &\textbf{0.980}$\pm$\textbf{0.00}     &\textbf{0.980}$\pm$\textbf{0.00}     &\textbf{0.980}$\pm$\textbf{0.00}     &0.980$\pm$0.00       \\
			
			\bottomrule
		\end{tabular}\vspace{-14pt}
	}
	
\end{table*}
\begin{itemize}
	\item \textbf{Twitter}: Twitter is a popular online social network that provides a platform for users to share their life with their online friends. Lots of the tweets written by users in Twitter are location-related along with timestamp. Our dataset includes $4,893$ users and $9,490,707$ tweets. $257,248$ locations appears along with tweets. Besides, the number of follow links between these users is $164,920$ in total.
	
	\item \textbf{Foursquare}: Foursquare is another famous social network allowing users to interact with friends online through multiple location-related services. Our dataset has $5,392$ users in Foursquare and $76,972$ friendship relationship among them. All these users have checked-in at $38,921$ different locations via $48,756$ tips. There are $3,282$ anchor links between Twitter and Foursquare in the dataset. 
\end{itemize}

\subsection{Experimental Settings}

\subsubsection{Experimental Setup}
In the experiments, we are able to acquire the set of anchor links between Foursquare and Twitter. The size of the set is $3,282$ which can be represented as $\mathcal{L}_+$. Based on the problem definition introduced in Section~\ref{subsec:problem_definition}, between the Foursquare and Twitter network, all the remaining non-existing anchor links can be represented as set $\mathcal{H}$. A proportion of non-anchor links are sampled randomly from $\mathcal{H} \setminus \mathcal{L}_+$ as negative set based on different negative-positive (NP) ratios $\theta$. NP-ratio $\theta$ in experiments ranges from 5 to 50 with the step length 5. The positive and negative link sets are divided into 10 folds. Among them, 1 fold will be used as the training set and the remaining 9 folds as the test set. In order to simulate the problem setting without enough labeled data, we further sample a small proportion of labeled instances from the 1-fold training set as the final training set. The sampling process is controlled by parameter sample-ratio $\gamma$, which takes values in \{10\%, 20\%, $\cdots$, 100\%\}. Here, $\gamma = 10\%$ denotes only $10\%$ of the 1-fold training set (i.e., only $1\%$ of the complete labeled data) is sampled in the final training set; while $\gamma = 100\%$ means all the instances in the 1-fold training set (i.e., $10\%$ of the labeled data) are used for training the model. In order to prevent unexpected impacts caused by data partitioning, we take 10 folds in turns to act as train set and the average metrics of 10 experiments are taken as the final results. We run the experiments on a Dell PowerEdge T630 Server with 2 20-core Intel CPUs and 256GB memory. The operating system is Ubuntu 16.04.3, and all codes are implemented in Python. 

\subsubsection{Comparison Methods}
The methods used in experiments are listed as following, we use them to verify 2 aspects of conclusions. One is the effectiveness of meta diagram based feature vector, and the other is the advantage of {\our}.
\begin{itemize}
	\item \textbf{\our}: {\our} is the model proposed in this paper which implements the learning process described in Section 3.4. Through a limited budget, we aim at selecting a good query set with the objective to improve the performance of {\our}. Two different versions of {\our} with budgets 50 and 100 are compared in the experiments.
	\item \textbf{\activerand}: In this method, we select the query set $\mathcal{U}_q$ in a random way in this method. The method is used to verify the effectiveness of the query set selection criteria used in {\our}.
	\item \textbf{\pusvm}: {\pusvm} extends the cardinality constrained link prediction model proposed in \cite{ZCZCY17} by incorporating the meta diagrams for feature extraction from aligned heterogeneous networks. ITER-MPMD is based on a PU (positive unlabeled) learning setting, without active query step. 
	\item \textbf{\svmmp}:  SVM is a classic supervised learning model. The feature vector used for building the SVM-MP model are extracted merely based on the meta paths.
	\item \textbf{\svmmpmd}: {\svmmpmd} is identical to {\svmmp} excepts it is built based on the features extracted with both meta paths and meta diagrams. Results comparison between {\svmmpmd} and {\svmmp} can verify the effectiveness of the meta diagram based features proposed in this paper. Meanwhile, comparison of {\svmmpmd} and {\pusvm} can also show that PU learning setting adopted in {\pusvm} is suitable for the network alignment problem.
\end{itemize}

\subsubsection{Evaluation Metrics}
We choose to use conventional evaluation metrics to measure the performance of different methods in experiments. The methods we test in experiments, including {\svmmp}, {\svmmpmd}, {\pusvm}, {\activerand}, and {\our}, can all output link prediction labels, and we will use F1, Recall, Precision and Accuracy as evaluation metrics. It should be noted that we need to query some labels in {\activerand} and {\our}. In other words, for the active-learning based methods, labels of these queried links are known already. In evaluation, we will remove these queried links from test set to maintain evaluation fairness between different comparison methods.

\subsection{Convergence and Scalability Analysis}
\begin{figure}[t]
	\centering
	\begin{minipage}[l]{0.7\columnwidth}
		\centering
		\includegraphics[width=\textwidth]{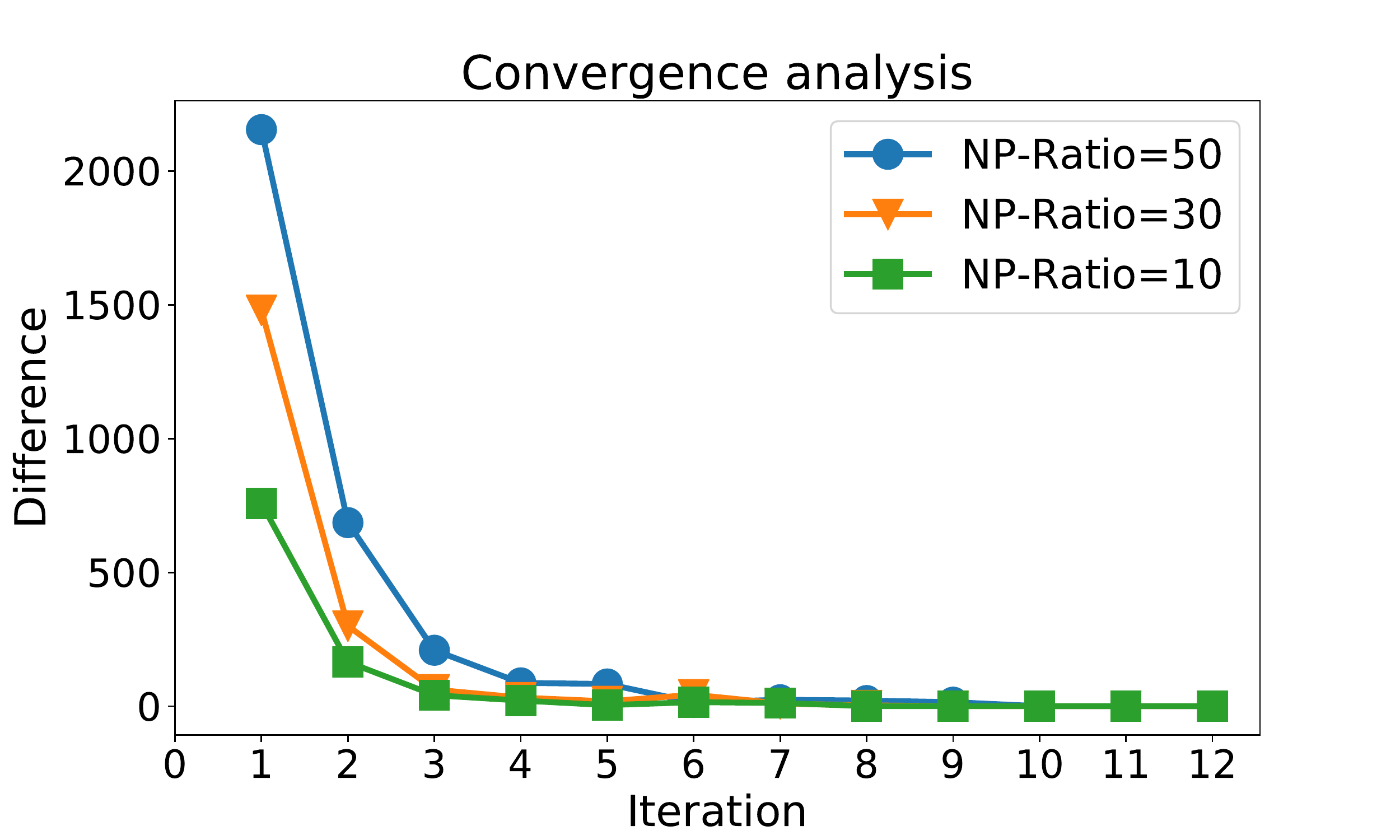}
	\end{minipage}
	\caption{Convergence analysis when sample-ratio=100\%.}\label{fig:convergence} 
	\vspace{-15pt}
\end{figure}
\begin{figure}[t]
	\centering
	\begin{minipage}[l]{0.7\columnwidth}
		\centering
		\includegraphics[width=\textwidth]{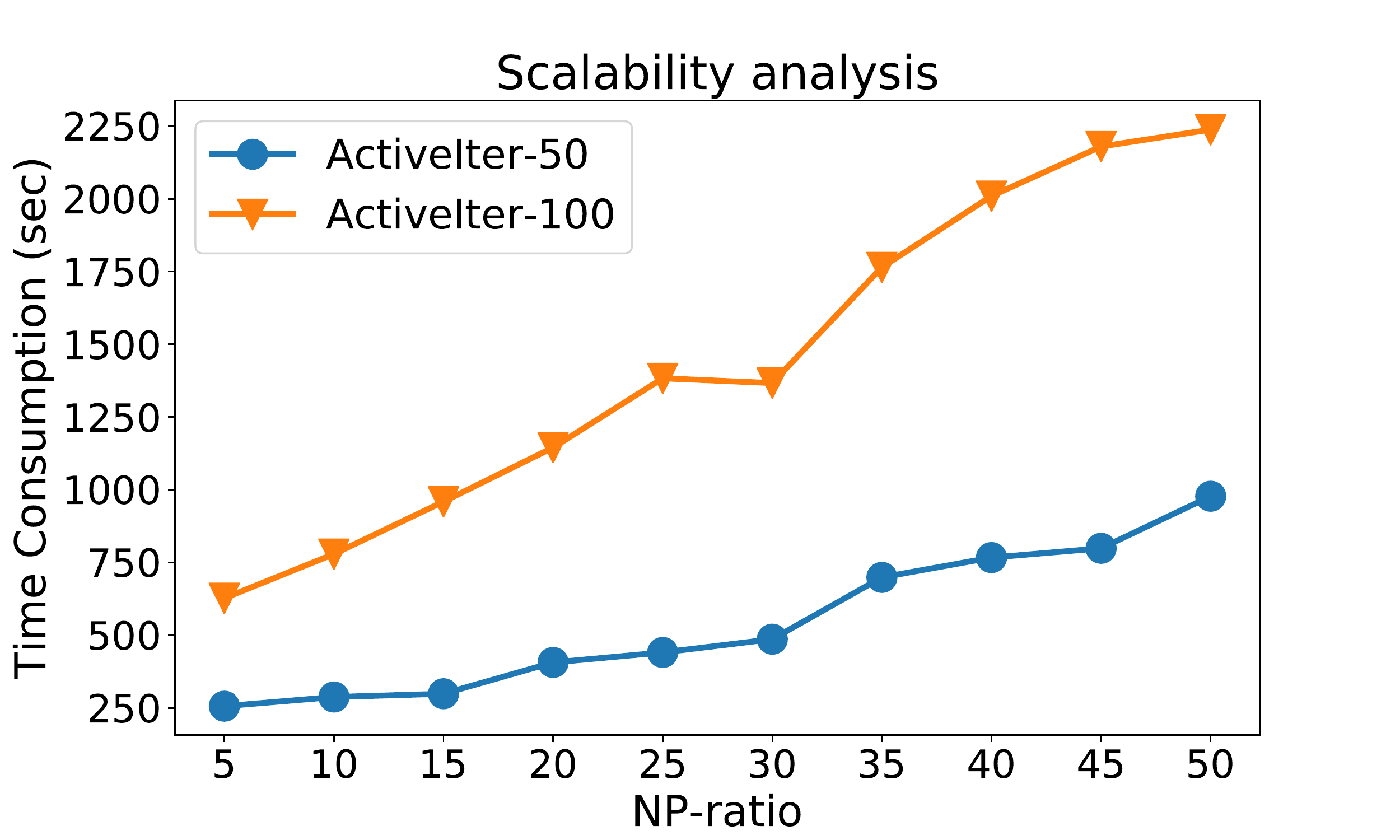}
	\end{minipage}
	\caption{Scalability analysis when sample-ratio=100\%.}\label{fig:scala} 
	\vspace{-15pt}
\end{figure}

In building the model {\our}, we propose to use the External Iteration Step (1) in the Section 3.4 essentially to learn both the model variable vector $\mb{w}$ and predict the anchor link label vector $\mb{y}$. In order to to show such an iteration step can convergence, in Figure~\ref{fig:convergence}, we show the label vector changes in each iteration. Here, the x axis denotes the iterations, and the y axis denotes the changes of vector $\mb{y}$ in sequential iterations $i$ and $i-1$, i.e., $\Delta \mb{y} = \left\| \mb{y}^i - \mb{y}^{i-1} \right\|_1$. According to Figure~\ref{fig:convergence}, we observe that the label vector of {\our} in the external iteration step can converge in less than $5$ iterations for different NP-ratios.\\
Figure~\ref{fig:scala} shows the near-linear scaling of {\our}'s running time in the data size. Here the X axis is the NP-ratio $\theta$, where the value of $\theta$ can represent the number of total links as we set before. The slopes indicate linear growth which shows the scalability of {\our}.

\subsection{Experimental Results with Analysis}
\begin{figure*}[t]
	\centering
	\subfigure[F1]{ \label{fig:link_prediction_analysis_1}
		\begin{minipage}[l]{0.45\columnwidth}
			\centering
			\includegraphics[width=1.1\textwidth]{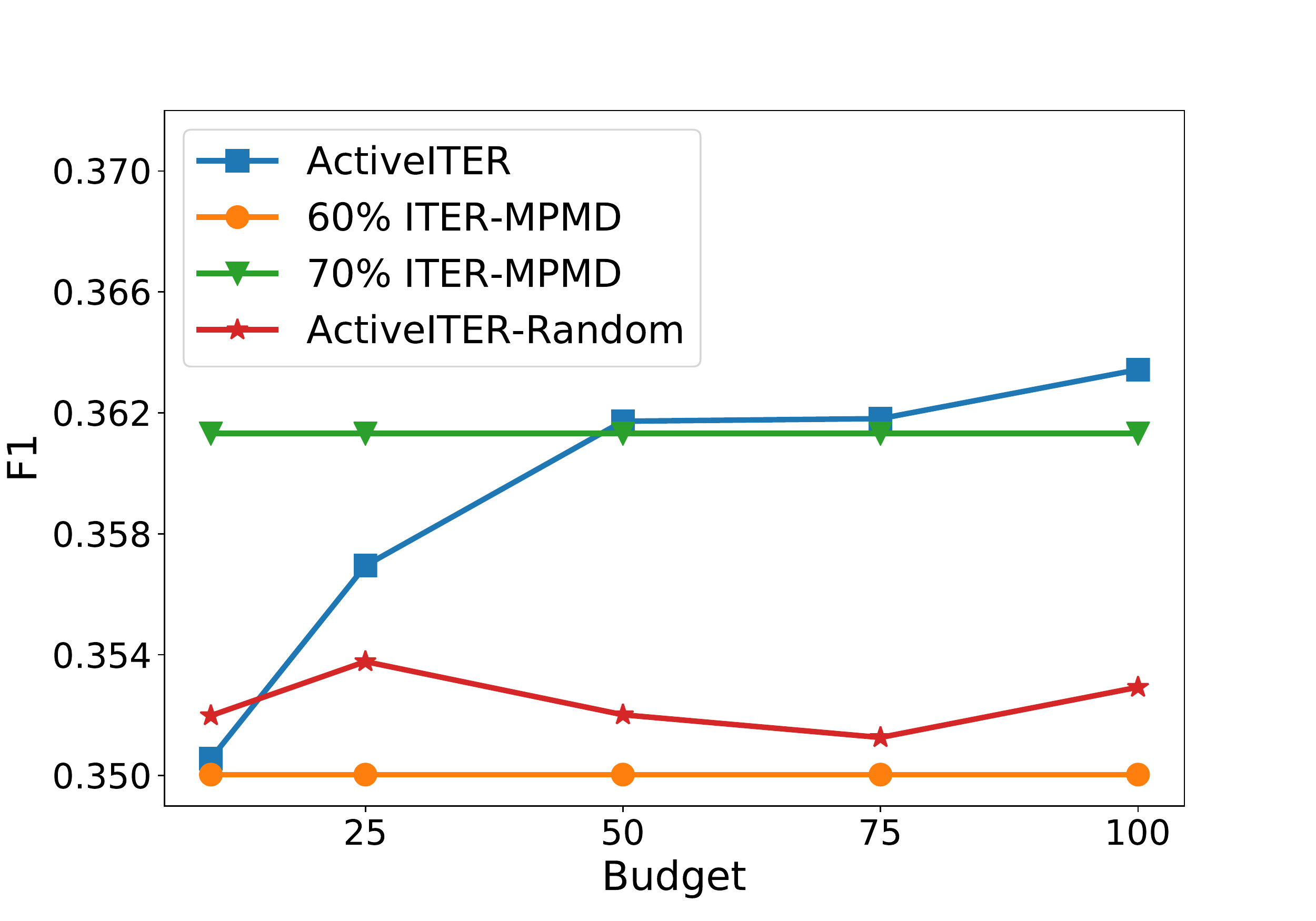}
		\end{minipage}
	}
	\subfigure[Recall]{\label{fig:link_prediction_parameter_analysis_2}
		\begin{minipage}[l]{0.45\columnwidth}
			\centering
			\includegraphics[width=1.1\textwidth]{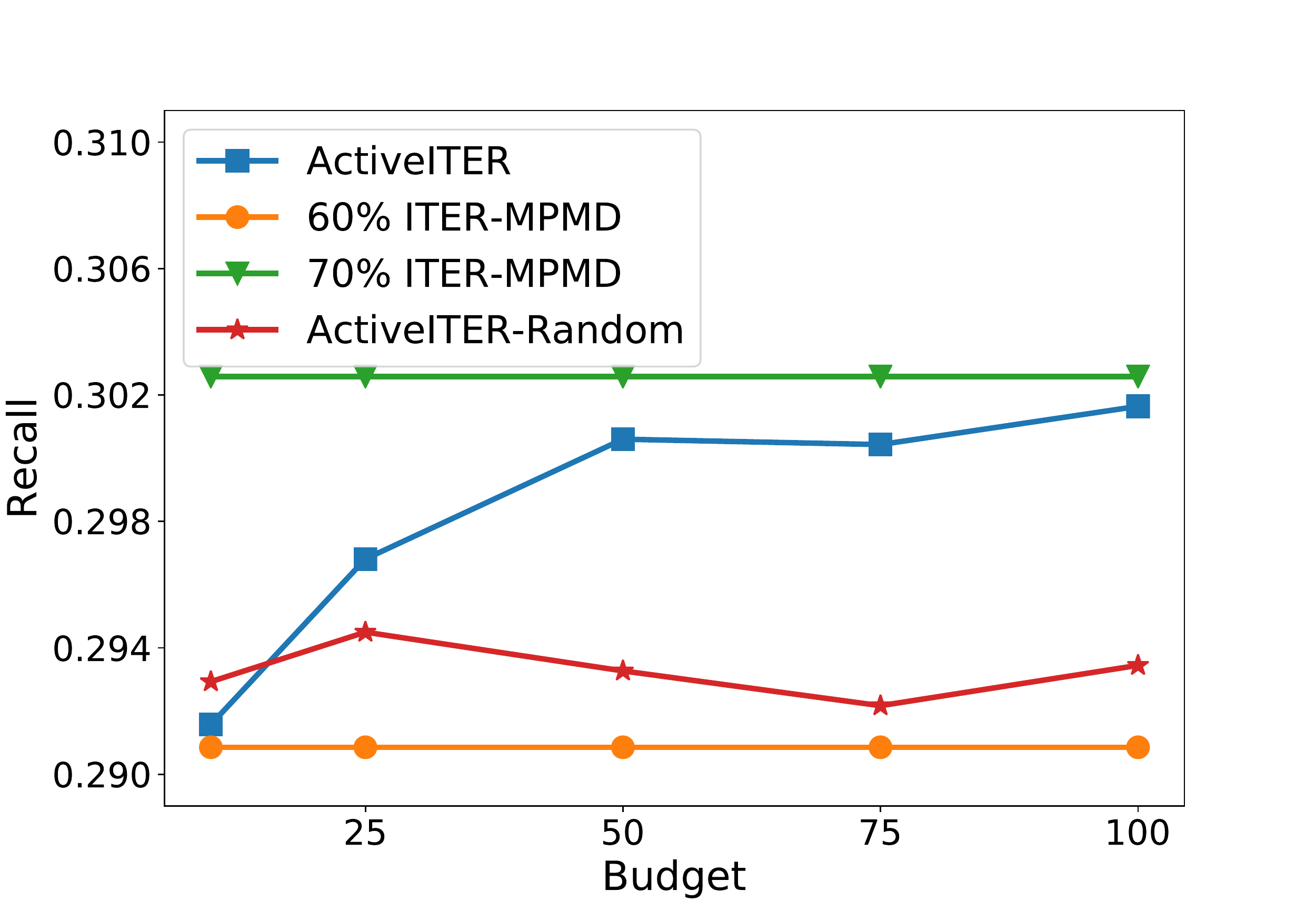}
		\end{minipage}
	}
	\subfigure[Precision]{ \label{fig:link_prediction_parameter_analysis_3}
		\begin{minipage}[l]{0.45\columnwidth}
			\centering
			\includegraphics[width=1.1\textwidth]{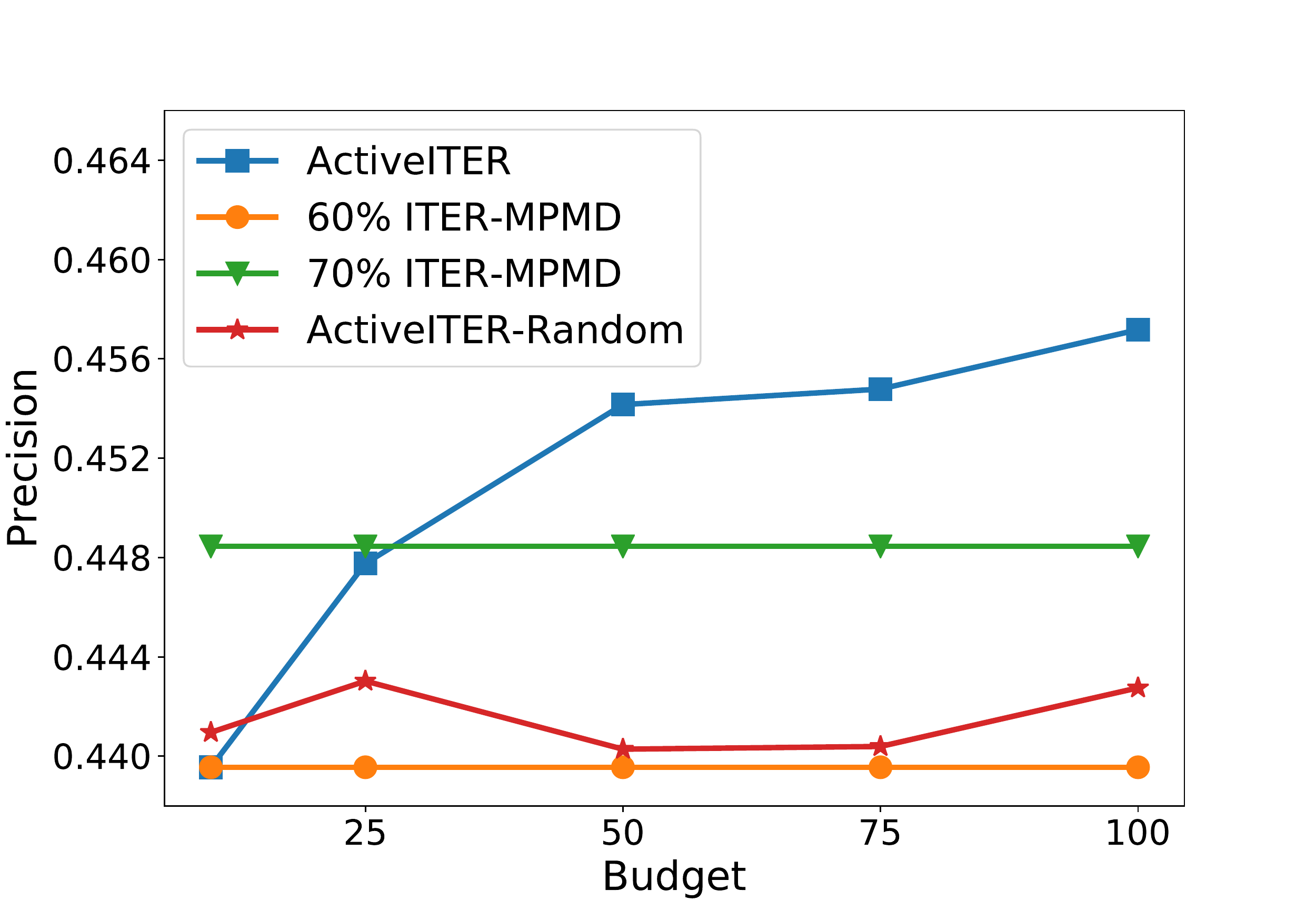}
		\end{minipage}
	}
	\subfigure[Accuracy]{ \label{fig:link_prediction_parameter_analysis_4}
		\begin{minipage}[l]{0.45\columnwidth}
			\centering
			\includegraphics[width=1.1\textwidth]{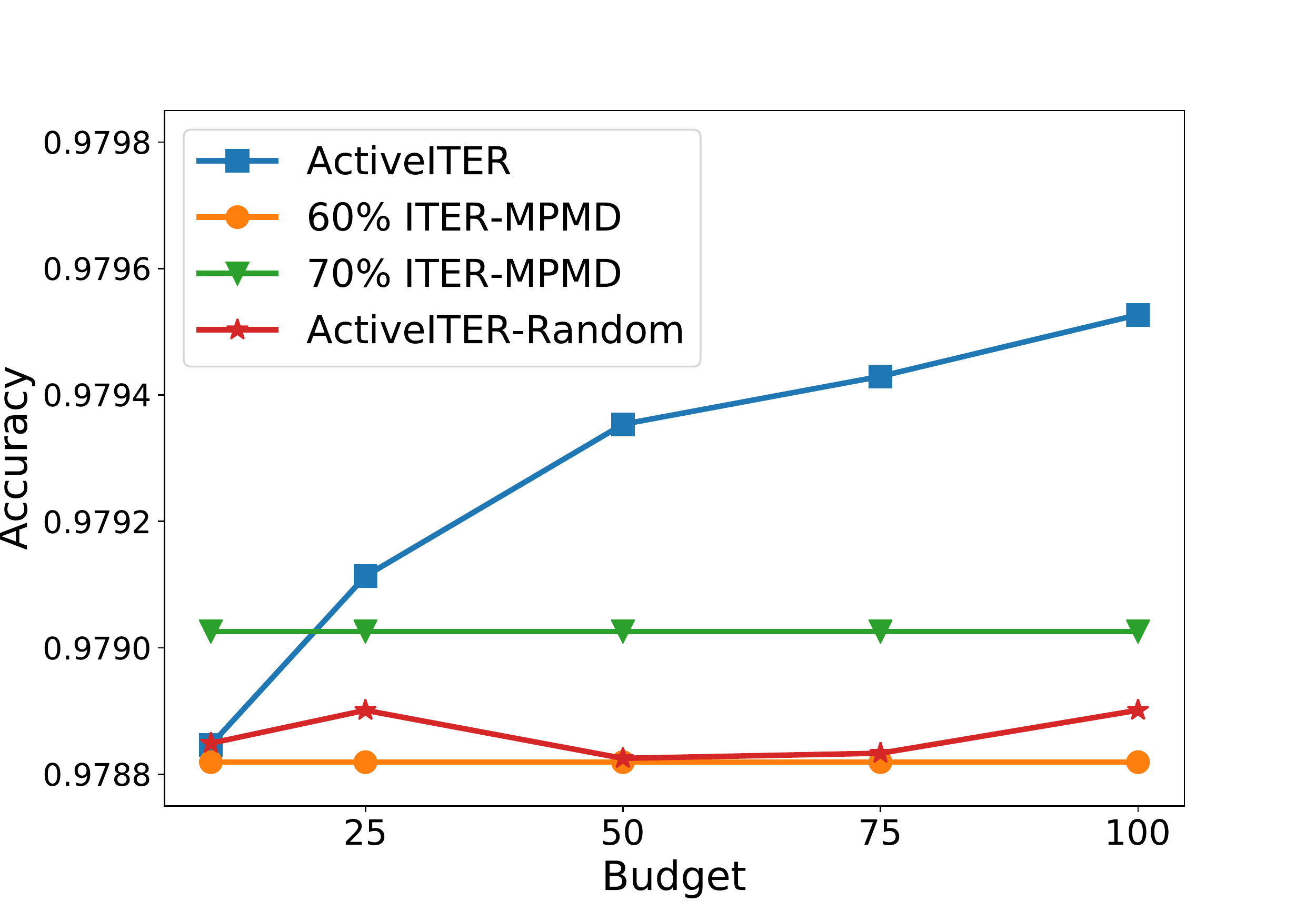}
		\end{minipage}
	}
	\caption{Performance Analysis under different budgets when $\theta=50$ and $\gamma=60\%$ for {\our} and {\activerand}.}\label{fig:link_prediction_analysis}
	\vspace{-10pt}
\end{figure*}

The experimental results acquired by different comparison methods are shown in Table~\ref{tab:main_result_fix_train_ratio} and Table~\ref{tab:main_result_fix_np_ratio} mainly. In Table~\ref{tab:main_result_fix_train_ratio}, Sample-ratio $\gamma$ is fixed as 60\%, and NP-ratio $\theta$ changes within  \{5, 10, $\cdots$, 50\}. The experimental results of these comparison methods are evaluated by the F1, Recall, Precision and Accuracy metrics respectively. Here, {\ourtenround} denotes {\our} with $50$ query budget, and {\ourtwentyround} has a query budget of value $100$. At first, we focus on the comparison between {\svmmp} and {\svmmpmd}. We can find {\svmmpmd} has a distinct advantage over {\svmmp} with $\theta \in \{5, 10, \cdots, 50\}$. Especially when $\theta$ is over 25, the Recall of {\svmmp} goes down to $0$, and it denotes {\svmmp} becomes ineffective in identifying the positive anchor links. However, {\svmmpmd} can still work in such a class imbalance scenario. There is only one exception in the table: when $\theta = 5$, the recall of {\svmmp} is better than {\svmmpmd}. We believe it is caused by very limited positive links and then conduct the supplementary experiment which samples the dataset another time and verifies the recall in $\theta=5$ is just an accident finally. Therefore, we can verify the effectiveness of the feature vector based on meta diagrams by the comparison of this set of experiments. Besides, the comparison between {\svmmpmd} and {\pusvm} demonstrates that {\pusvm} based on a PU learning setting provides a much better modeling for network alignment. However, we can find that the Accuracy of {\svmmpmd} is the highest when $\theta$ is over $45$. Here, we should remind when $\theta$ is high enough, {\svmmpmd} can not predict positive links correctly which can be found from its Recall. Therefore, in such a class-imbalance setting, Accuracy cannot work well in evaluating the comparison methods performance any more. 

Meanwhile, by comparing {\pusvm} with {\activerandom}, we can discover the metrics obtained by {\activerandom} can even be worse than {\pusvm} in some cases. In other words, querying labels in a random way will not contribute to the improvement of the prediction result. From the results, we are also able to observe that {\ourtenround} outperforms {\activerandom} consistently for $\theta \in \{5, 10, \cdots, 50\}$. In addition, the comparison between {\ourtenround} and {\ourtwentyround} shows the budget value may have an impact on the performance of {\our}, whose sensitivity analysis is available in Section~\ref{sec:parameter}. 

In Table~\ref{tab:main_result_fix_np_ratio}, we fix $\theta$ as $50$ and change the sample-ratio $\gamma$ with values in  \{10\%, 20\%, $\cdots$, 100\%\}. From Table~\ref{tab:main_result_fix_np_ratio}, we can confirm conclusions verified from Table~\ref{tab:main_result_fix_train_ratio} are still valid firstly. Furthermore, we can make comparison between {\ourtwentyround} with certain $\gamma$ and {\pusvm} with $\gamma+10\%$, When $\theta = 50$, the size of training set will increase by $1,670$, if $\gamma$ increases by $10\%$. Between these two methods, besides the $\gamma$  percentage of training instances shared by both methods, ITER-MPMD uses additional $1,670$ training instances, while {\ourtwentyround} merely queries for additional $100$ instances. According to the results, in most of the cases, {\ourtwentyround} with far less training data can still outperform ITER-MPMD with great advantages. For example, when $\gamma = 80\%$, {\ourtwentyround} has metrics that $F1=0.3978$, $Precision=0.4913$, $Recall=0.3343$ and $Accuracy=0.9804$. We use {\pusvm} which $\gamma = 90\%$ as a comparison. $F1$, $Precision$, $Recall$ and $Accuracy$ achieved by {\pusvm} are $0.3875$, $0.4755$, $0.3270$ and $0.9797$ respectively. In other words, {\our} can get better performance with around $5\%$ cost in labeling links compared with {\pusvm}.

\subsection{Parameter Analysis}\label{sec:parameter}

The effects of the parameter budget $b$ on the performance of {\our} will be analyzed in this part. From Figure~\ref{fig:link_prediction_analysis}, we can observe that {\our} can achieve better prediction results consistently along with querying critical labels continuously, but {\activerand} can not improve prediction output with random labels. This result shows that when $b$ rises, {\our} is accompanied by better results in all metrics including F1, Precision. Recall and Accuracy. Meanwhile, this performance improvement is continuous and significant because when the $b$ changes within  \{10, 25, 50, 75, 100\}, the improvement of performance does not slow down. After $b$ exceeds $50$, three key metrics including F1, Precision and Accuracy have been higher than {\pusvm} which has $1,670$ more labeled links in the training set. According to the analysis results, with far less (less than 100 additional) training instances, method {\our} proposed in this paper based on active learning can achieve comparable and even better results than the non-active method {\pusvm} with 1,670 extra training instances.


\section{Related Work} \label{sec:related_work}

Network alignment problem is an important research problem, which has been studied in various areas, e.g., protein-protein-interaction network alignment in bioinformatics \cite{KBS08, LLBSB09, SXB07}, chemical compound matching in chemistry \cite{SHL08}, data schemas matching data warehouse \cite{MGR02}, ontology alignment web semantics \cite{DMDH04}, graph matching in combinatorial mathematics \cite{MH14}, and figure matching and merging in computer vision \cite{CFSV04, BGGSW09}. Network alignment is an important problem for bioinformatics. By studying the cross-species variations of biological networks, network alignment problem can be applied to predict conserved functional modules \cite{SSKKMUSKI05} and infer the functions of proteins \cite{PSBLB11}. Graemlin \cite{FNSMB06} conducts pairwise network alignment by maximizing an objective function based on a set of learned parameters. Some works have been done on aligning multiple network in bioinformatics. IsoRank proposed in \cite{SXB08} can align multiple networks greedily based on the pairwise node similarity scores calculated with spectral graph theory. IsoRankN \cite{LLBSB09} further extends IsoRank by exploiting a spectral clustering scheme.

Similarity measure based on heterogeneous networks has been widely studied. Sun introduces the concept of \textit{meta path-based similarity} in \textit{PathSim} \cite{SHYYW11}, where a \textit{meta path} is a path consisting of a sequence of relations. However, the \textit{meta path} suffers from two disadvantages. On one hand, \textit{meta path} cannot describe rich semantics effectively. On the other hand, once numerious \textit{meta paths} are defined, it's challenging to assemble them. Some methods to resolve these  deficiencies are proposed later. \textit{Meta structure} \cite{HZCSML16} applys meta-graph to similarity measure problem, but entities are constrained to be of the same type. Zhao \cite{ZYLSL17} proposes the concept of \textit{meta graph} and extends the idea to recommendation problems which require that entities belong to different types. However, \textit{meta structure} and \textit{meta graph} are proposed for single non-attribute networks. In our \textit{inter-network meta diagram} definition, not only regular node types but also attribute types are involved, and it can be applied to the similarity measure across networks.

For online social networks, network alignment provides an effective way for information fusion across multiple information sources. In the social network alignment model building, the anchor links are very expensive to label manually, and achieving a large-sized anchor link training set can be extremely challenging. In the case when no training data is available, via inferring the potential anchor user mappings across networks, Zhang et al. have introduced an unsupervised network alignment models for multiple social networks in \cite{ZY15} and an unsupervised network concurrent alignment model via multiple shared information entities simultaneously in \cite{ZY16}. However, pre-labeled anchor links can provide necessary information for understanding the patterns of aligned user pairs in their information distribution, which lead to the better performance than the unsupervised alignment models. Therefore, in \cite{ZY15_ijcai, ZCZCY17}, Zhang et al. propose to study the network alignment problem based on the PU learning setting.

Active learning is an effective method for network alignment in the face of lacking labeled links which has been previous studied by \cite{CS13, EAE17}. The query strategies proposed by Cort\'{e}s and Serratosa \cite{CS13} return a probability matrix for different alignment choices which makes the quantification of network alignment straightforward. However, this kind of strategies totally ignore the \textit{one-to-one cardinality constraint} existing in online social networks. Therefore, we provide an innovative query strategy considering \textit{one-to-one cardinality constraint} in {\our}. Malmi \cite{EAE17} proposes two relative-query strategies \textsc{TopMatching} and \textsc{GibbsMatching} instead of focusing on absolute-query. However, it may not be less challenging for experts to make comparative judgements in online social networks, because the quantity of cantidates corresponding to one node will be huge.  

Across the aligned networks, various application problems have been studied. Cross-site heterogeneous link prediction problems are studied by Zhang et al. \cite{ZKY14} by transferring links across partially aligned networks. Besides link prediction problems, Jin and Zhang et al. proposes to partition multiple large-scale social networks simultaneously in \cite{ZY15-2}. The problem of information diffusion across partially aligned networks is studied by Zhan et al. in \cite{ZZWYX15}, where the traditional LT diffusion model is extended to the multiple heterogeneous information setting. Shi et al. give a comprehensive survey about the existing works on heterogeneous information networks in \cite{SLZSY15}, which includes a section talking about network information fusion works and related application problems in detail.

\vspace{-5pt}
\section{Conclusion}\label{sec:conclusion}

In this paper, we study the {\problem} problem and propose an active learning model {\our} based on meta diagrams to solve this problem. Meta diagrams can be extracted from the network to constitute heterogeneous features. In our experiments, we verify the effectiveness of meta diagram based feature vectors at first. In the active learning model {\our}, we propose an innovative query strategy in the selection process to in order to query for the optimal unlabeled links. Extensive experiments conducted on two real-world networks \textit{Foursquare} and \textit{Twitter} demonstrate that {\our} has very outstanding performance compared with the state-of-the-art baseline methods. {\our} only needs a small-size training set to build up initially and can outperform the other non-active models with much less training instances.

\section{Acknowledgements}
This work is partially supported by FSU and by NSF through grant IIS-1763365.

\balance
\bibliographystyle{plain}
\bibliography{reference}


\end{document}